\documentclass[a4paper,12pt]{article}
\usepackage{fullpage}
\usepackage{amssymb,amsmath}
\usepackage{xspace}
\usepackage{graphicx}
\usepackage{amssymb}
\usepackage{amsmath}
\usepackage{multicol}

\usepackage{caption}
\usepackage{subcaption}
\usepackage{float}
\usepackage[percent]{overpic}
\usepackage{tikz}
\usepackage{hyperref}

\newcommand{\pa}{\partial}

\newcommand{\Tr}{\mathrm{Tr}}

\newcommand{\hog}{Hedgehog\xspace}
\newcommand{\sk}{Skyrmion\xspace}
\newcommand{\sks}{Skyrmions\xspace}
\newcommand{\bx}{\mbox{\boldmath $x$}}
\newcommand{\bX}{\mbox{\boldmath $X$}}
\newcommand{\bJ}{\mbox{\boldmath $J$}}

\newcommand{\bI}{\mbox{\boldmath $I$}}

\newcommand{\fg}{figure\xspace}

\newcommand{\fgs}{figures\xspace}

\newcommand{\btau}{\mbox{\boldmath $\tau$}}
\newcommand{\bpi}{\mbox{\boldmath $\pi$}}

\RequirePackage{snapshot}

\begin{document}

\title{\vskip -70pt
\begin{flushright}
{\normalsize DAMTP-2015-18} \\
\end{flushright}
\vskip 50pt
{\bf \Large \bf Scattering of Nucleons in the Classical Skyrme Model}
 \vskip 30pt
\author{
David Foster$^\dagger$ and Nicholas S. Manton$^\star$\\[40pt]
{\em \normalsize $^\dagger$ School of Physics, HH Wills Physics 
Laboratory,}\\[0pt] 
{\em \normalsize University of Bristol, Tyndall Avenue, 
Bristol BS8 1TL, U.K.}\\ 
{\normalsize Email: \quad dave.foster@bristol.ac.uk}\\[5pt]
{\em \normalsize $^\star$ Department of Applied Mathematics and 
Theoretical Physics,}\\[0pt] 
{\em \normalsize University of Cambridge, Wilberforce Road, 
Cambridge CB3 0WA, U.K.}\\ 
{\normalsize Email: \quad N.S.Manton@damtp.cam.ac.uk}
}}

\maketitle
\vskip 20pt

\maketitle
\begin{abstract}
Classically spinning $B=1$ \sks can be regarded as approximations 
to nucleons with quantised spin. Here, we investigate nucleon-nucleon 
scattering through numerical collisions of spinning \sks. 
We identify the dineutron/diproton and dibaryon 
short-lived resonance states, and also the stable deuteron state. 
Our simulations lead to predictions for the polarisation states 
occurring in right angle scattering.

\end{abstract}
\section{Introduction}

The Skyrme model \cite{Skyrme:1961vq} is a nonlinear theory of pion 
fields that has topological soliton solutions. 
Encouragingly, Witten identified the Skyrme model as a low energy 
effective model of QCD \cite{Witten:1983tw, Witten:1983tx}. The conserved 
topological charge is interpreted as the baryon number $B$, and the minimal 
energy static solutions for each integer $B$ are called \sks. They can be
treated as rigid bodies, free to rotate in space and also in isospace
(the three-dimensional space of the pion fields). When the rotational 
motion is quantised, the Skyrmions are models for nucleons and
nuclei. In particular the $B=1$ \sks, quantised with spin and isospin 
half, model protons and neutrons. 

A full quantum mechanical treatment of the interaction of two $B=1$
\sks is difficult. Each Skyrmion has three position coordinates and
three orientational coordinates, so two-Skyrmion dynamics involves
twelve coordinates \cite{AtiMan:1993}. A truncation to ten coordinates 
has been useful for modelling the deuteron bound state 
\cite{Leese:1994hb}, but no proper discussion of
two-nucleon scattering is possible with this truncation. 


An alternative is a classical approach, using the idea that a classically
spinning Skyrmion is a reasonable model of a nucleon
\cite{GisPar,Manton:2011mi}. The classical angular velocity of the Skyrmion is fixed to match the quantised spin and isospin of the nucleon. Scattering of non-spinning Skyrmions was first performed using an axially symmetric ansatz \cite{Verbaarschot:1986rj}, the first full-field simulation was
reported in \cite{Allder:1987kq}, and multi-charge \sk scattering was considered in \cite{Battye:1996nt}. The scattering of spinning Skyrmions representing nucleons was considered by Gisiger and Paranjape
\cite{GisPar}, and they obtained analytical formulae for the scattering angle for large impact parameters. There has also been some recent work on multi-charge, purely isospinning \sks \cite{Battye:2014qva}.

A systematic, numerical investigation of Skyrmion scattering without spin, at moderate and small impact parameters \cite{Foster:2014vca}, confirmed that matter is exchanged between the Skyrmions when the
Skyrmions are close, and also showed that a substantial rotation of the \sk's orientation can occur. Here, we numerically investigate the scattering of two classically spinning $B=1$ Skyrmions to model
proton-proton, neutron-neutron and proton-neutron scattering at various impact parameters, including head-on collisions. We are particularly interested in the change in the polarisation states
of the nucleons when they scatter, and also in finding evidence,
using our classical approximation, for the deuteron and for the
known two-nucleon resonance states.

The format of this paper is as follows. We first introduce the Skyrme
model and its calibration. Then we discuss how to identify classically
spinning \sks as protons or neutrons. The next section presents the
results of our numerical scattering of Skyrmions, and our
identification of the dineutron/diproton resonances, the deuteron
bound state, and the excited dibaryon state. The concluding section 
summarises and discusses our results.

\section{The Skyrme model}
The Skyrme model is defined by the Lagrangian density
\begin{align}
\mathcal{L}= -\frac{F_\pi^2}{16} \Tr(R_\mu R^\mu)
+ \frac{1}{32e^2}\Tr([R_\mu , R_\nu][R^\mu , R^\nu])
+ \frac{m_\pi^2 F_\pi^2}{8}\Tr(U-I_2) \,,
\label{Lag}
\end{align}
where the Skyrme field $U(t,\bx)$ is an $\mbox{SU}(2)$-valued scalar
and $R_\mu=\pa_\mu U U^\dag$ is its $\mbox{su}(2)$-valued current. 
$F_\pi$, $e$ and $m_\pi$ are parameters, which are fixed by comparison with 
experimental data. Their values will be discussed later. It is convenient 
for us to work in dimensionless Skyrme units. One Skyrme length unit
corresponds to $\frac{2}{eF_\pi}$ in inverse MeV, and one Skyrme
energy unit corresponds to $\frac{F_\pi}{4e}$ MeV. Conversion of
inverse MeV to fm is as usual through $\hbar = 197.3$ MeV fm. The 
dimensionless pion mass in Skyrme units is
\begin{align}
m=\left(\frac{2}{eF_\pi}\right)m_\pi \,. \label{pion mass}
\end{align}
In Skyrme units, the energy of a static field is
\begin{equation}
E = \int \left( -\frac{1}{2} \mbox{Tr}(R_i R_i) - \frac{1}{16}
\mbox{Tr}([R_i,R_j][R_i,R_j]) + m^2\mbox{Tr}(I_{2}-U) \right)\, d^3x
\,. \nonumber \label{SM}
\end{equation}
It is often convenient, especially in numerical simulations, to 
express $U$ in terms of a triplet of pion fields 
$\bpi=(\pi_1,\pi_2,\pi_3)$ and an additional
auxiliary field $\sigma$ as
\begin{align}
U(t,\bx)=\sigma(t,\bx) \, I_2 + i \bpi(t,\bx)\cdot \btau \,,
\end{align}
where $\btau=(\tau_1,\tau_2,\tau_3)$ are the three Pauli matrices,
with the constraint $\sigma^2+\bpi\cdot\bpi =1$ so that $U \in
\mbox{SU}(2)$.  

At fixed time, $U(t,\bx)$ is a map $U:\mathbb{R}^3\to
\mbox{SU}(2)$, where $U\to I_2$ at spatial infinity. This boundary
condition compactifies $\mathbb{R}^3\cup\{\infty\}$ to $S^3$. The
group manifold of $\mbox{SU}(2)$ is $S^3$, so a finite energy
configuration $U$ extends to a map $U:S^3 \to S^3$, and
then belongs to a class of $\pi_3(S^3) = \mathbb{Z}$ indexed by an 
integer $B \in \mathbb{Z}$, called the baryon number. $B$ is also 
the degree of the map $U$ which can be explicitly calculated as
\begin{equation}
B \equiv \int \mathcal{B}(\bx) \, d^3x = -\frac{1}{24\pi^2} \int
\varepsilon_{ijk}\Tr(R_i R_j R_k) \, d^3x \,, \label{baryon-density}
\end{equation}
where $\mathcal{B}(\bx)$ is the baryon density. \sks are the 
static field configurations of minimal energy for each value of 
$B$. In the \fgs we plot level-sets of baryon density $\mathcal{B}(\bx)$.

\begin{figure}[H] 
       \centering
       \begin{subfigure}[b]{0.3\textwidth}
               \centering
    \includegraphics[width=\textwidth]{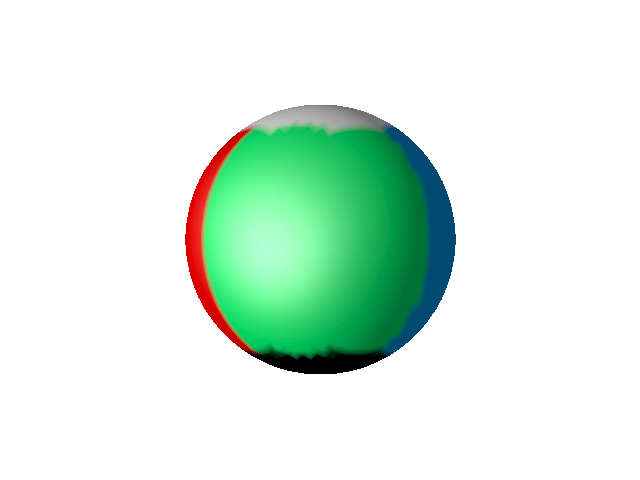}
               \caption*{Side}
    
       \end{subfigure}%
        ~ 
        \begin{subfigure}[b]{0.3\textwidth}
                \centering \includegraphics[width=\textwidth]{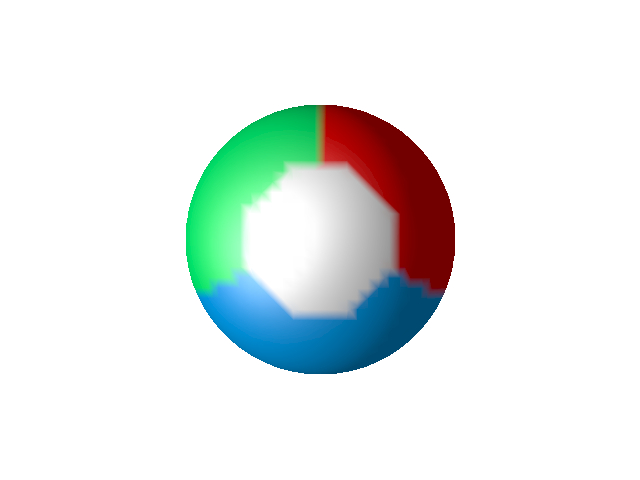}
                \caption*{Top}
         
        \end{subfigure}
        \caption{$B=1$ \hog \sk.}
        \label{B1Skyrme}
\end{figure}

It is well known that the $B=1$ \sk can be found using 
the so-called Hedgehog ansatz \cite{Skyrme:1961vq}
\begin{align} \label{hog-ansatz}
U_H(\bx)=\cos f(r) \, I_2+i\sin f(r) \, \hat{\bx} \cdot \btau \,,
\end{align}
where $r=|\bx|$ and $\hat{\bx}$ is the radial unit vector. $f(r)$ is a 
real radial profile function satisfying $f(0)=\pi, f(\infty)=0$.
This produces a solution with rotationally symmetric energy and baryon
density. Figure \ref{B1Skyrme} shows a $B=1$ \hog \sk coloured as
in \cite{Manton:2011mi}. The centres of the white and black regions
are where $\pi_1^2+\pi_2^2=0$ and  $\pi_3 >0$, $\pi_3 <0$ respectively. 
The centres of the red, green and blue regions are where $\pi_3=0$ and
$\tan^{-1}\left(\frac{\pi_2}{\pi_1}\right)=0,\frac{2\pi}{3},\frac{4\pi}{3}$
respectively. This is the colouring scheme used throughout this paper.

An important feature of the \hog ansatz is that an isorotation, $U(\bx) \to
AU(\bx)A^\dag$, which acts as a rotation among the pion fields, is
equivalent to a rotation in space, $U(\bx) \to U(D(A)\bx)$, where 
$D(A)_{ij}=\frac{1}{2}\Tr(\tau_i A \tau_j A^\dag)$ is the standard
$\mbox{SO}(3)$ rotation matrix corresponding to the $\mbox{SU}(2)$ matrix $A$.
This can be understood in terms of the coloured \sks,
where a spatial rotation can be `undone' by a reordering of colours,
i.e. an isorotation of the pion fields. A further use of the colouring 
is that it shows when two \hog \sks are in the attractive channel 
\cite{Schroers:1993yk}. This is when the colours on the nearest, 
facing sides of two \sks match.
 
\begin{figure}[H] 
                \centering \includegraphics[width=0.4\textwidth]{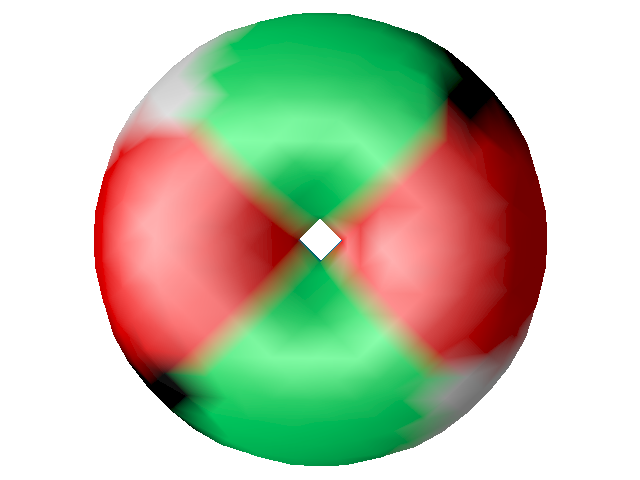}
        \caption{$B=2$ \sk.}
        \label{B2 Skyrme}
\end{figure}

The $B=2$ \sk plays an important role in the dynamics of two $B=1$
Skyrmions. This \sk is toroidal, and is shown in figure \ref{B2 Skyrme}.

To relate the Skyrme model to nuclear physics, \sks should be
quantised. As the \sks are energy minima they can be treated as
rigid bodies rotating in space and isospace. For general \sks,
semi-classical quantisation is the quantisation of the time-dependent
space and isospace rotations $A_2$ and $A_1$ in the rigid-body ansatz
\begin{align}
U(t,\bx)=A_1(t)U_0(D(A_2(t))\bx)A_1(t)^\dag \,, 
\end{align}
where $U_0(\bx)$ is a static solution. $A_1(t)$ and $A_2(t)$ are
$\mbox{SU}(2)$ matrices, and $D(A_2(t))$ is, as before, the
$\mbox{SO}(3)$ matrix corresponding to $A_2(t)$. The spherical
symmetry of the \hog ansatz \eqref{hog-ansatz} implies that the 
rotational motion of the $B=1$ Skyrmion only depends on the
combined $\mbox{SU}(2)$ matrix $A(t) = A_1(t)A_2(t)$.

Under rigid rotation the $B=1$ \sk has body-fixed angular momentum
$L_i=\lambda(-a_i+b_i)$ and body-fixed isospin angular momentum
$K_i=\lambda(a_i-b_i)$, where $a_j=-i \, \Tr(\tau_j A_1^\dag \dot{A_1})$
is the angular velocity in isospace, $b_j=i \, \Tr(\tau_j \dot{A_2}
A_2^\dag)$ is the angular velocity in space and $\lambda$ is the
moment of inertia \cite{Battye:2009ad}. From these one obtains the
space-fixed angular momentum $J_i=-D(A_2)^T_{ij} L_j$ and the
``space-fixed'' isospin angular momentum $I_i=-D(A_1)_{ij} K_j$.
The space-fixed angular momenta correspond to physical spin and
isospin, and these can be expressed purely in terms of $A(t)$ and 
its time derivative. Nucleons have both spin and isospin half. 
A proton is an isospin-up state, and a neutron is an isospin-down state. 

Due to the spherical symmetry of the \hog ansatz, $L_i+K_i=0 \ 
(i=1,2,3)$. Something similar is true for the $B=2$ toroidal \sk where 
$L_3+2K_3=0$ \cite{Braaten:1988cc}. We shall make use of these properties
later. For an in-depth discussion of \sk quantisation we point the
reader to \cite{Adkins:1983hy,Braaten:1988cc,Manko:2007pr,Battye:2009ad}.

\subsection{\sk calibration}
For the model to be relevant to nuclear physics it requires
calibration. Adkins and Nappi \cite{Adkins:1983hy} calibrated the
quantised $B=1$ \sk against the masses of the spin $\frac{1}{2}$ 
nucleons, the spin $\frac{3}{2}$ delta resonances, and the pions, finding
$F_\pi = 108$ MeV, $e=4.84$ and $m_\pi =138$ MeV (so $m = 0.526$). This
calibration assumes that rigid rotation is a good approximation, but
the surface fields of the delta are rotating close to the speed of
light and pion radiation is very strong. Therefore, for the delta, the
approximation is not reliable and this calibration is not valid for
higher charge Skyrmions. Recently, Manton and Lau calibrated the 
Skyrme model against the states of Carbon-12 \cite{Lau:2014baa} and 
found $m=0.7$ to be optimal. For the $B=1$ Skyrmion this leads
to an energy (\sk mass) $M=159.89$, $\lambda=55.89$ and an rms matter radius
$\langle r^2 \rangle^{1/2}=1.006$ in Skyrme units. The calibration requires 
$F_\pi = 117.5$ MeV and $e=3.93$. Planck's constant in Skyrme units is
always $\hbar=2e^2$ so for this calibration $\hbar=30.8$. These are
the values we use for our simulations and throughout this paper:
\begin{equation}
m=0.7~\mbox{and}~\hbar=30.8 \,. \nonumber
\end{equation}

The resulting nucleon mass is the 
static $B=1$ \sk energy $M$ plus the spin-energy contribution, $M + \frac{3}{8 \lambda} \hbar^2$, which in physical units is
\begin{align}
M_{\rm N}=M \frac{F_\pi}{4e} + \frac{3}{8 \lambda}e^3F_\pi\,.
\end{align}
With our calibration
$M_{\rm N}=1243$ MeV, about $30\%$ larger than the physical nucleon
mass $M_{\rm N}=939$ MeV. Also the rms matter radius is $\langle r^2 \rangle^{1/2}=0.87$ fm. 

The $B=2$ toroidal Skyrmion can also be quantised as 
a rigid body. The two lowest-energy states are a spin 1, isospin 0 state
representing the deuteron, and a spin 0, isospin 1 state representing
the low-energy resonance known as the diproton/dineutron.
With our calibration, and following the quantisation
procedure in \cite{Manko:2007pr}, we find the mass for the
deuteron to be $2324$ MeV, which is similarly larger than the
experimental value of $1876$ MeV. We also find the rms matter radius
to be $1.08$ fm, which is less than half the experimentally found
rms radius of the deuteron $\sim2.14$ fm, showing that rigid-body
quantisation of the static $B=2$ \sk gives a too tightly bound representation
of the loosely bound deuteron. A better representation of the deuteron in
the Skyrme model was obtained in \cite{Leese:1994hb}. Using a Yang--Mills 
instanton ansatz for the Skyrme field \cite{Atiyah:1989dq}, an
additional, radial degree of freedom could be included, allowing the two 
$B=1$ Skyrmions to separate. In the deuteron state, this radial degree 
of freedom oscillates with the zero point motion of a vibrational
ground state. We will see later that modest oscillations of the
radial degree of freedom are generated when spinning Skyrmions collide
classically. 

Although this calibration is not optimal for the proton, neutron or
deuteron, we have found that the processes discussed later are not
sensitive to small changes of calibration.

\section{Spinning \sks as protons or neutrons}

It was discovered by Adkins, Nappi and Witten \cite{Adkins:1983ya} 
that the properly normalised wavefunctions for the proton and neutron 
in spin-up and spin-down states, along the $z$-axis, are
\begin{align}
p^\uparrow &= \frac{1}{\pi}(a_1+ia_2), ~~~ p^\downarrow = -\frac{i}{\pi}(a_0-ia_3), \\
n^\uparrow &= \frac{i}{\pi}(a_0+ia_3), ~~~ n^\downarrow = -\frac{1}{\pi}(a_1-ia_2),
\end{align}
where $A=a_0I_2+ia_i\tau_i$ is the $\mbox{SU}(2)$ matrix that controls
the $B=1$ \sk's orientation and $a_0^2+a_1^2+a_2^2+a_3^2=1$. It has
been proposed that these wavefunctions can be used
to identify protons and neutrons as classically spinning Skyrmions
\cite{GisPar, Manton:2011mi}. This is because each wavefunction has maximal
magnitude on a great circle in $\mbox{SU}(2)$ and motion 
along the great circle is equivalent to a $4\pi$ rotation of the \sk in space.  
The phase of the wavefunction changes by $2\pi$ around the circle, so
the spin is $\frac{1}{2}$.

The wavefunctions of $p^\downarrow$ and $n^\uparrow$ are maximal when
the Hedgehog Skyrmion (initially in its standard orientation) is simply 
rotated about the $z$-axis, whereas the 
wavefunctions of $p^\uparrow$ and $n^\downarrow$ are maximal when the
Skyrmion is flipped over, and then rotated about the $z$-axis. 

A \hog \sk spinning with angular frequency $\omega$ about the
$z$-axis is obtained by acting with the isorotation matrix 
$A(t)=\exp(\frac{i\omega t}{2}\tau_3)
=\cos(\frac{\omega t}{2})+i\sin(\frac{\omega t}{2})\tau_3$.
This isospinning \sk is simultaneously spinning about 
its white-black axis in space. It has spin and isospin projections
$J_3=\lambda\omega$, $I_3=-\lambda \omega$, which are opposite. 
An anticlockwise spin, $\omega>0$, corresponds to an $n^\uparrow$
state and a clockwise spin, $\omega<0$, corresponds to a $p^\downarrow$ state. 

The isorotation matrix $A(t)(i\tau_2)$ flips the \sk by $\pi$ about 
the $y$-axis, and rotates it with angular frequency $\omega$ about 
the $z$-axis. Such an isospinning Skyrmion has $J_3=-\lambda\omega$ and
$I_3=-\lambda \omega$, so the spin and isospin projections are now equal. 
An anticlockwise spin, $\omega>0$, corresponds to an
$n^\downarrow$ state and a clockwise spin, $\omega<0$, corresponds 
to a $p^\uparrow$ state. 

\begin{figure}[H]
\centering
\includegraphics[width=0.55\textwidth]{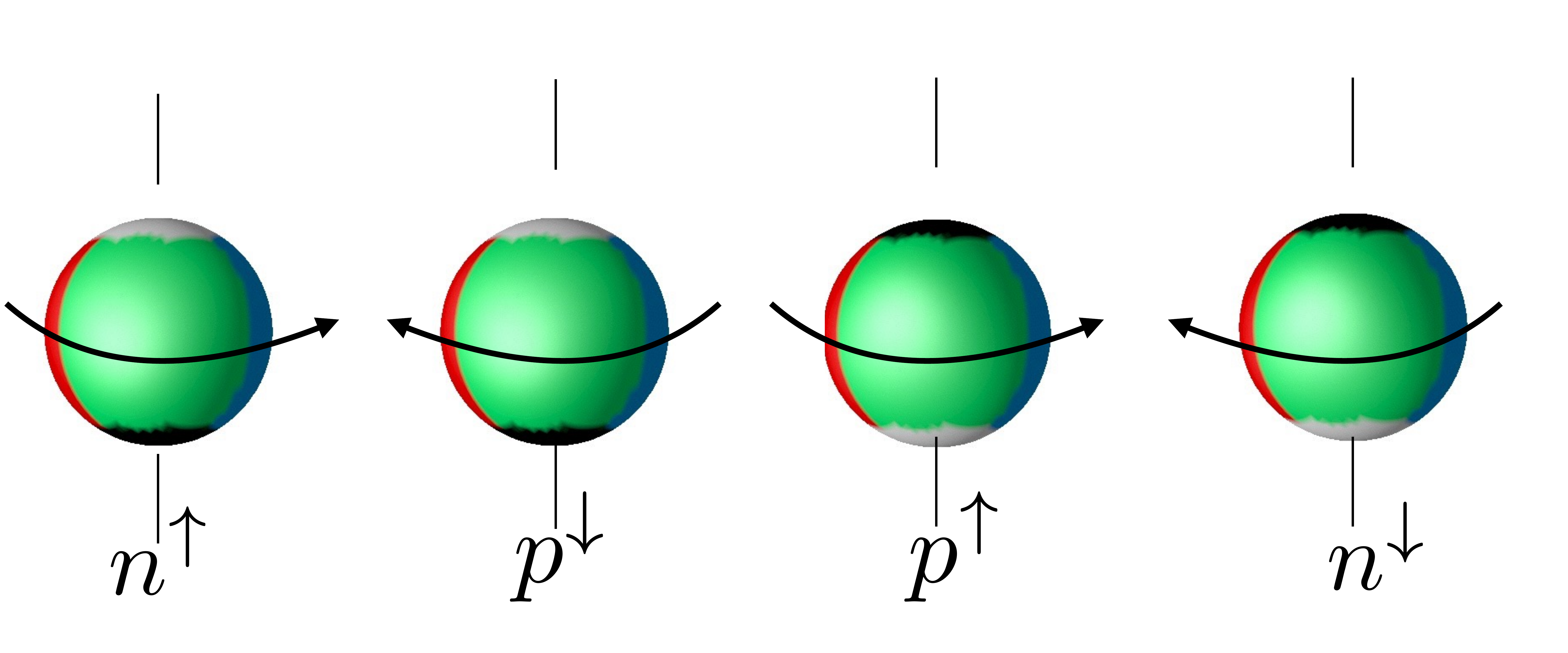}
\caption{Identifying \sks as neutrons and protons.}
\label{spinning skyrme}
\end{figure}

The replacement $A(t) \to A(t)(i\tau_2)$ produces the spin 
flips $n^\uparrow \to n^\downarrow$ and $p^\downarrow \to p^\uparrow$  
because $D_{33}(A) \to -D_{33}(A)$. This leads to the
identification that a \sk spinning anticlockwise about its white-black
axis is always a neutron and a \sk spinning clockwise about the same 
axis is always a proton, as shown in figure \ref{spinning skyrme}. This 
is regardless of the orientation of the axis in space.

To classically model a nucleon whose projected 
spin has magnitude $\frac{1}{2}$ we require $\lambda|\omega| 
= \frac{1}{2} \hbar$, so
\begin{equation}
|\omega| = \frac{\hbar}{2\lambda} = 0.28 \,, \nonumber
\end{equation}
as $\hbar = 30.8$ and $\lambda = 55.89$. We use $|\omega|=0.28$
throughout this paper. With this angular velocity, the surface fields
of the \sk are not rotating close to the speed of light. 

There are classical models of the two lowest-energy quantum
states of the $B=2$ toroidal Skyrmion too. In the deuteron state, the
torus spins around an axis orthogonal to the symmetry axis of the torus, 
with spin 1. The colour orientation is arbitrary and unchanging in time, as
the deuteron has isospin zero. In the dineutron/diproton state the
torus is coloured as in figure \ref{B2 Skyrme}, and the colours circulate
steadily, so that the isospin is 1. The black and white regions on 
the circular edge of the torus are fixed. The torus is not spatially 
rotating, as the dineutron/diproton has spin zero. The torus does not 
rotate about its symmetry axis in either state. If it did, the constraint 
$L_3 + 2K_3 = 0$ would imply that the state had non-zero spin and
non-zero isospin.

\section{\sk scattering and its interpretation}

\subsection{Spinning \sk collisions}
We are interested in classically replicating nucleon scattering,
namely proton-proton, neutron-neutron and proton-neutron
scattering. To do this we identify protons and
neutrons as spinning \sks, as discussed above. Instead of
colliding non-spinning \sks as in some previous work
\cite{Foster:2014vca}, we numerically collide spinning \sks. We 
initiate a collision by forming two configurations $U_1$ and $U_2$ out of 
spinning and translated \hog \sks,
\begin{align}
U_1(t,\bx)=U_H\big(D_1(t)(\bx-\bX)\big), \nonumber \\
U_2(t,\bx)=U_H\big(D_2(t)(\bx+\bX)\big),
\end{align}
where $D_1(t)$ and $D_2(t)$ are two $\mbox{SO}(3)$ time-dependent
rotation matrices and $\pm\bX$ are the locations of the \sks. We
choose $\bX=(b,Y,0)$, where $Y$ is large compared with the \sk radius. 
Then, using the product ansatz, we create an initial configuration of two
\sks boosted towards each other, at velocities $\pm\mbox{v}$, by
\begin{align}
U(t,\bx)=U_1\big(x,\gamma(y+\mbox{v}t),z\big)U_2\big(x,
\gamma(y-\mbox{v}t),z\big) \,,
\end{align}
where $\gamma =1/\sqrt{1-\mbox{v}^2}$ is the usual Lorentz boost factor.
 
Each \sk is initially moving parallel to the $y$-axis. $b$ is the impact
parameter chosen to give the desired orbital angular momentum
$l\hbar=2M\mbox{v}b$, where $M$ is the Skyrmion mass. We define 
transverse polarisation to be where the \sk's white-black axis is 
aligned parallel to the $z$-axis and linear polarisation to be where the 
\sk's white-black axis is aligned parallel to the $y$-axis.

We numerically evolve these initial configurations using the field 
equations obtained from the Skyrme Lagrangian density (\ref{Lag}). We 
use a finite difference leap-frog numerical algorithm on a suitably large
lattice. Leap-frog was chosen because it is a symplectic integrator.

All of the simulations are presented as videos at \url{http://www.bristoltheory.org/people/david.foster/skyrmevideos/}; we urge the reader to view these videos.

\subsection{The Dineutron/Diproton}

There are no physical bound states in the $2$-proton or $2$-neutron
systems. However, a $2$-proton low energy resonance is observed in
proton-proton collisions and is called the diproton. It has spin $\bJ=0$
and isospin $\bI=1$. The dineutron is the corresponding
$2$-neutron resonance. Neutron-neutron scattering is less easy to achieve 
than proton-proton scattering, but the dineutron resonance is observed as a 
decay product of the neutron-rich nucleus Beryllium-16, $^{16}$Be 
\cite{PhysRevLett.108.102501}, where it is energetically favourable 
for $^{16}$Be to decay to $^{14}$Be by simultaneously ejecting two neutrons. 
The trajectories of the two detected neutrons trace back to the same
point, where they form a
state of lifetime $\sim10^{-22}$ s. The $^{16}$Be and $^{14}$Be ground
states are both $\bJ=0$, hence the dineutron is a $\bJ=0$ state. 
The relative energy of the two neutrons is found to be $\approx
0.25$ MeV. So, assuming no other radiation, the neutrons separate with
initial speed $\mbox{v}=0.02$.
 
\begin{figure}[H]
    \begin{subfigure}[b]{0.3\textwidth}
     \centering \begin{overpic}[width=\textwidth]{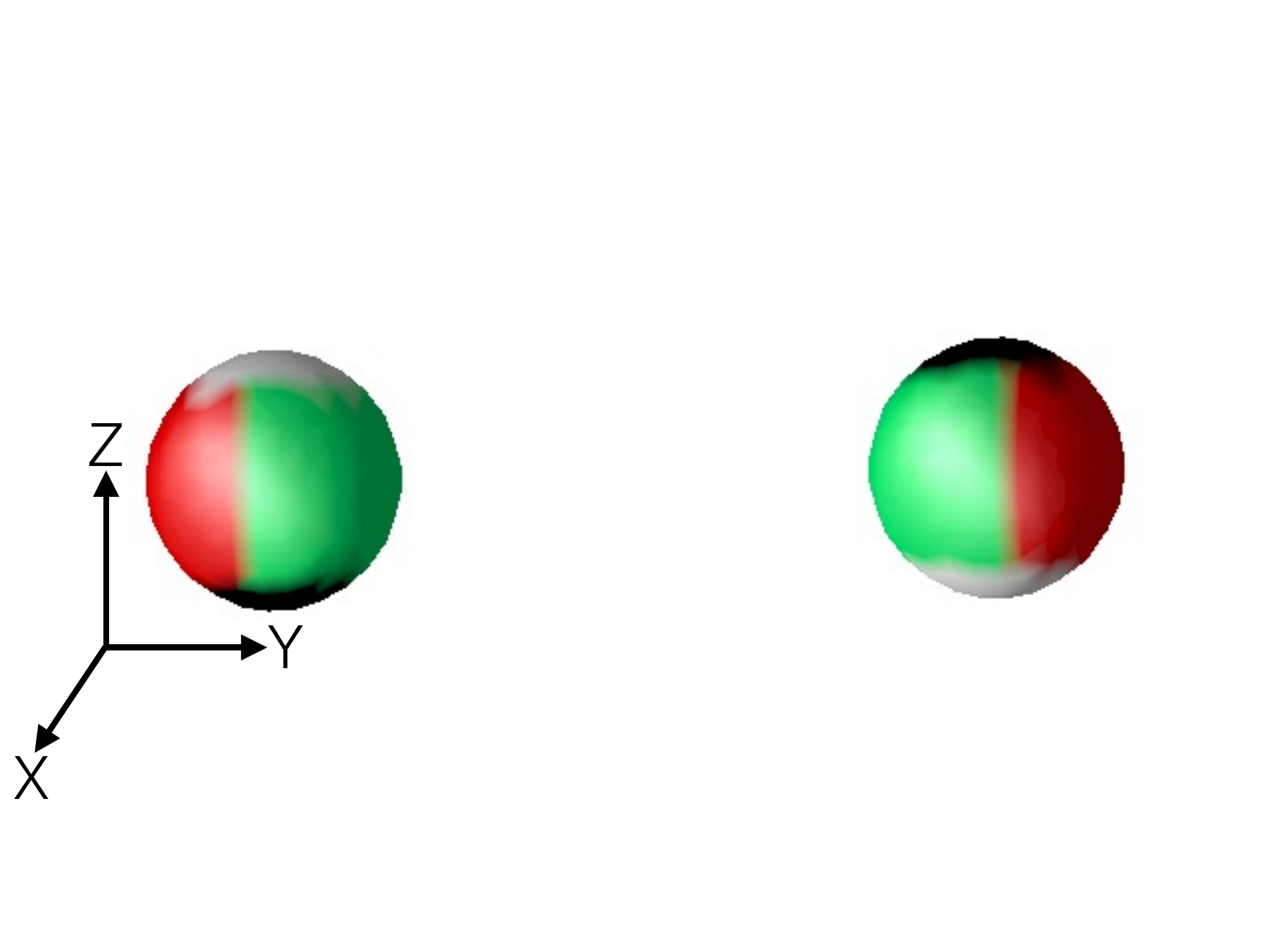}
 \put (10,63) {Transverse}
  \put (10,55) {polarisation}
\end{overpic}
       \caption*{$t=t_0$}
        \end{subfigure} 
        ~
               \begin{subfigure}[b]{0.3\textwidth}
                \centering \includegraphics[width=\textwidth]{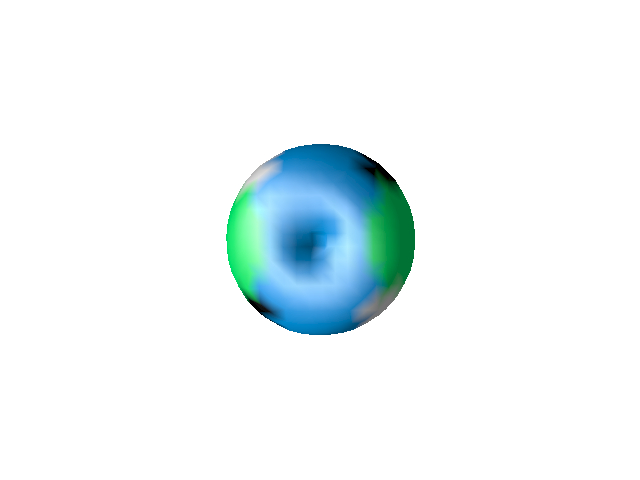}
                \caption*{$t=t_1$}
     
        \end{subfigure}
        ~
          \begin{subfigure}[b]{0.3\textwidth}
                \centering \includegraphics[width=\textwidth]{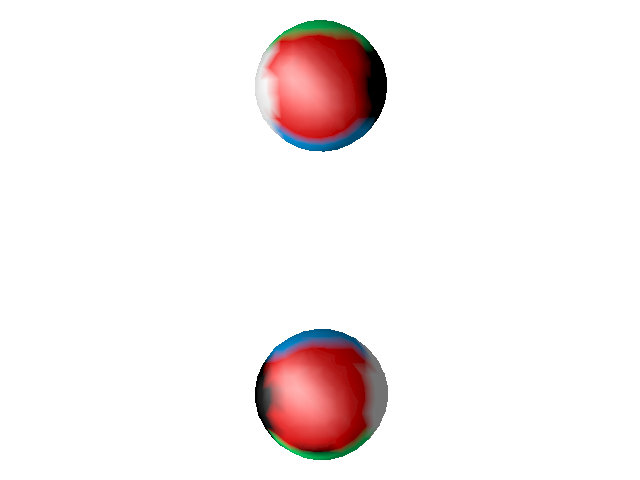}
                \caption*{$t=t_2$}
           
        \end{subfigure}
        \\ \\ \\
\begin{subfigure}[b]{0.3\textwidth}
               \centering
       \begin{overpic}[width=\textwidth]{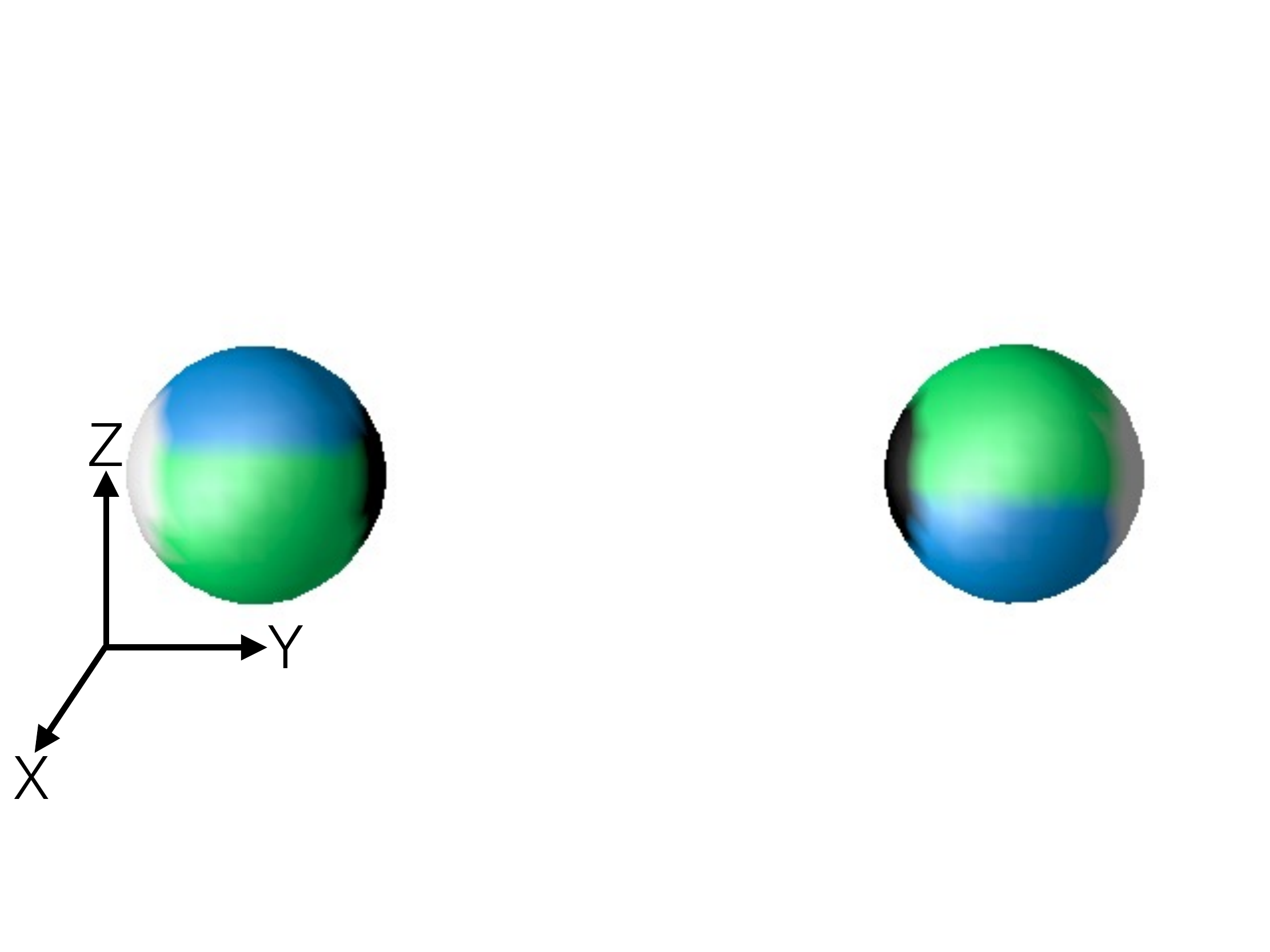}
 \put (10,63) {Linear}
 \put (10,55) {polarisation}
\end{overpic}
               \caption*{$t=t_0$}
        	       \end{subfigure}%
        ~ 
        \begin{subfigure}[b]{0.3\textwidth}
                \centering \includegraphics[width=\textwidth]{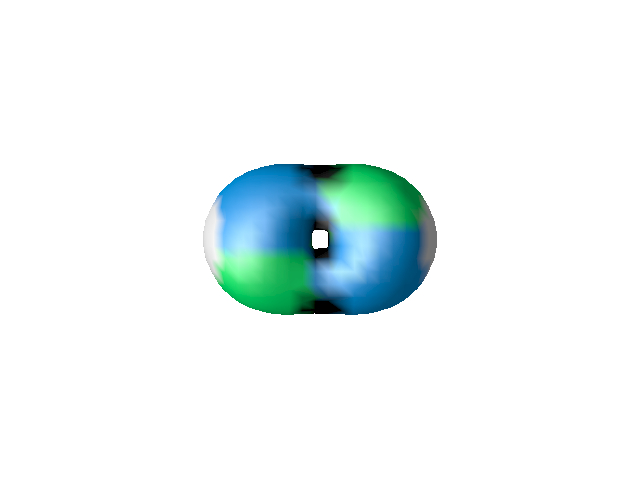}
                \caption*{$t=t_1$}     
        \end{subfigure}
        ~
          \begin{subfigure}[b]{0.3\textwidth}
                \centering \includegraphics[width=\textwidth]{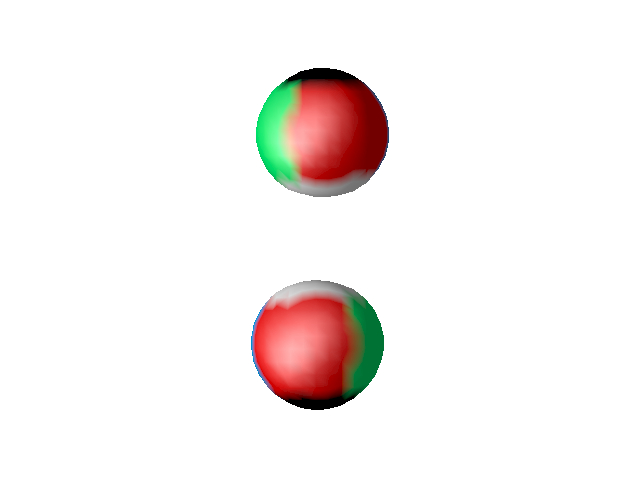}
                \caption*{$t=t_2$}
           
        \end{subfigure}
        \caption{Level-sets of baryon density $\mathcal{B}(\bx)$ for
          the $\bJ=0$ scattering of two neutrons. Three
          time-snapshots are shown here. The transverse polarisation 
          scattering and the linear polarisation scattering videos are shown online.}
        \label{j0I1-scattering}
\end{figure}

There is no significant electrostatic repulsion between two neutrons,
and since we have not included Coulomb effects in our Skyrme model, our
simulations of Skyrmion scattering are likely to better 
model the dineutron rather than the diproton. So we collide two \sks 
which are spinning as anti-aligned neutrons (illustrated in 
figure \ref{spinning skyrme}), giving a $\bJ=0$, $\bI=1, I_3 = -1$ state, 
with initial speeds $\mbox{v}=0.02$. The \sks are either transversely 
polarised (parallel to the $z$-axis) or linearly polarised. Both 
types of scattering are shown in \fg \ref{j0I1-scattering}.

It is seen that no bound state is
produced, but the $B=2$ torus configuration is present for a short 
time. This torus forms in the $(y,z)$-plane and it does not spin, 
but it has isospin because the colouring continues to circulate at all 
times. It has the quantum numbers $\bJ=0$ and $\bI=1, I_3 = -1$ which is
consistent with the short-lived dineutron state observed in the decay
of $^{16}$Be. For both polarisations, the black and white regions 
alternate around the circular edge of the torus. 

The scattering shown in figure \ref{j0I1-scattering} is for a head-on
collision of two neutrons. For both polarisations, the outgoing 
Skyrmions are moving at right angles to the incoming Skyrmions, and
they are spinning as neutrons too, because of isospin conservation. 
Similar right angle scattering was observed for non-spinning Skyrmions in
\cite{Foster:2014vca}. 

In the case of transverse polarisation the outgoing direction 
of motion is precisely the direction of the
initial polarisation axis. There is another interesting phenomenon: the 
polarisation axis, the white-black axis, jumps by $90$ degrees 
during the scattering, from being parallel to the $z$-axis to being
parallel to the $y$-axis. This is possible because of the way the 
Skyrmions merge and break up, such that each outgoing Skyrmion is a 
combination of half of each incoming Skyrmion. 

From our numerical simulations we can observe the sense of the 
polarisation, before and after the collision. For example, 
the neutron incoming from the negative $y$-direction (the left) has its 
polarisation in the positive $z$-direction, and the neutron that 
is outgoing in the positive $z$-direction has its polarisation 
in the negative $y$-direction. The result can be stated in a more 
invariant way, both for $2$-neutron and $2$-proton right angle 
scattering. Each particle, incoming or outgoing, has a momentum 
${\bf p}$ and classical spin ${\bf s}$. The two incoming 
particles and the two outgoing particles all have the same value for 
the vector ${\bf p} \times {\bf s}$, so the scattering is accompanied 
by spin rotation.

In the case of linear polarisation the outgoing direction of motion can be any
direction at right angles to the $y$-axis, and is determined by the
precise initial colour orientations. The scattering shown in figure 
\ref{j0I1-scattering} has been selected to be from the $y$-direction 
to the $z$-direction. Again there is a $90$ degree jump 
in the polarisation axis, so that after
scattering the neutrons are still linearly polarised. There is however
a flip in the sense of the polarisation. In our example, for both incoming 
neutrons the polarisation is outwards, but for the outgoing neutrons it is
inwards. Helicity, ${\bf p} \cdot {\bf s}$, is therefore the same for
incoming and outgoing particles, and this result is more general, holding 
in the diproton case too.

Although the scattering here is classical, the results suggest that
quantised $2$-neutron and $2$-proton scattering depends strongly on
polarisations. For Skyrmions transversely polarised in the $z$-direction, 
the outgoing motion breaks the axial symmetry around the $y$-axis, 
because the outgoing particle motion is concentrated in the $z$-direction.
Also, the outgoing polarisations are completely determined.
It would be interesting to compare these classical simulations with
physical proton-proton scattering in which the polarisations of both incoming 
particles and both outgoing particles were measured.

\subsection{Deuteron formation} 

The deuteron is a stable, two-baryon bound state comprised of a proton and a
neutron. It has been understood as the quantised ground state of the
$B=2$ \sk \cite{Braaten:1988cc,{Leese:1994hb},Manko:2007pr} with spin 
$\bJ = 1$ and isospin $\bI = 0$. 

Here we are interested in classically modelling deuteron formation, 
so we numerically collide two spinning Skyrmions, representing a 
proton and neutron with transverse polarisation parallel to the 
$z$-axis. The deuteron total angular momentum projected along the 
$z$-direction has three contributions, the individual spins $J_{\rm p}$ and 
$J_{\rm n}$ of the proton and neutron, which are $\pm\frac{1}{2}$, 
and the orbital angular momentum, an integer $l$, combining to 
give $J_3=J_{\rm p}+J_{\rm n}+l$ \cite{krane1987introductory}. Quantum models 
predict that the $l=0$ state dominates, and the deuteron spin is 
accounted for by the proton and neutron spins being aligned. We 
can replicate the deuteron state by colliding 
two \sks head-on, where one \sk is spinning as a neutron with 
$J_{\rm n}=\frac{1}{2}$ and the other \sk has the opposite orientation 
of the white-black axis and is spinning as a proton with 
$J_{\rm p}=\frac{1}{2}$. The impact parameter is zero, so $l=0$.

The deuteron has slightly lower energy than a proton plus a neutron, so 
we cannot use a simple energetic 
argument to choose an initial collision speed. Whatever the initial speed,
some energy needs to be dissipated in order to truly produce a
deuteron. It is therefore a considerable surprise that our collisions
do produce a configuration similar to what is expected for a
deuteron in a classical Skyrmion model.

\begin{figure}[H]
       \centering
           
          \begin{subfigure}[b]{0.3\textwidth}
               \centering
    \includegraphics[width=\textwidth]{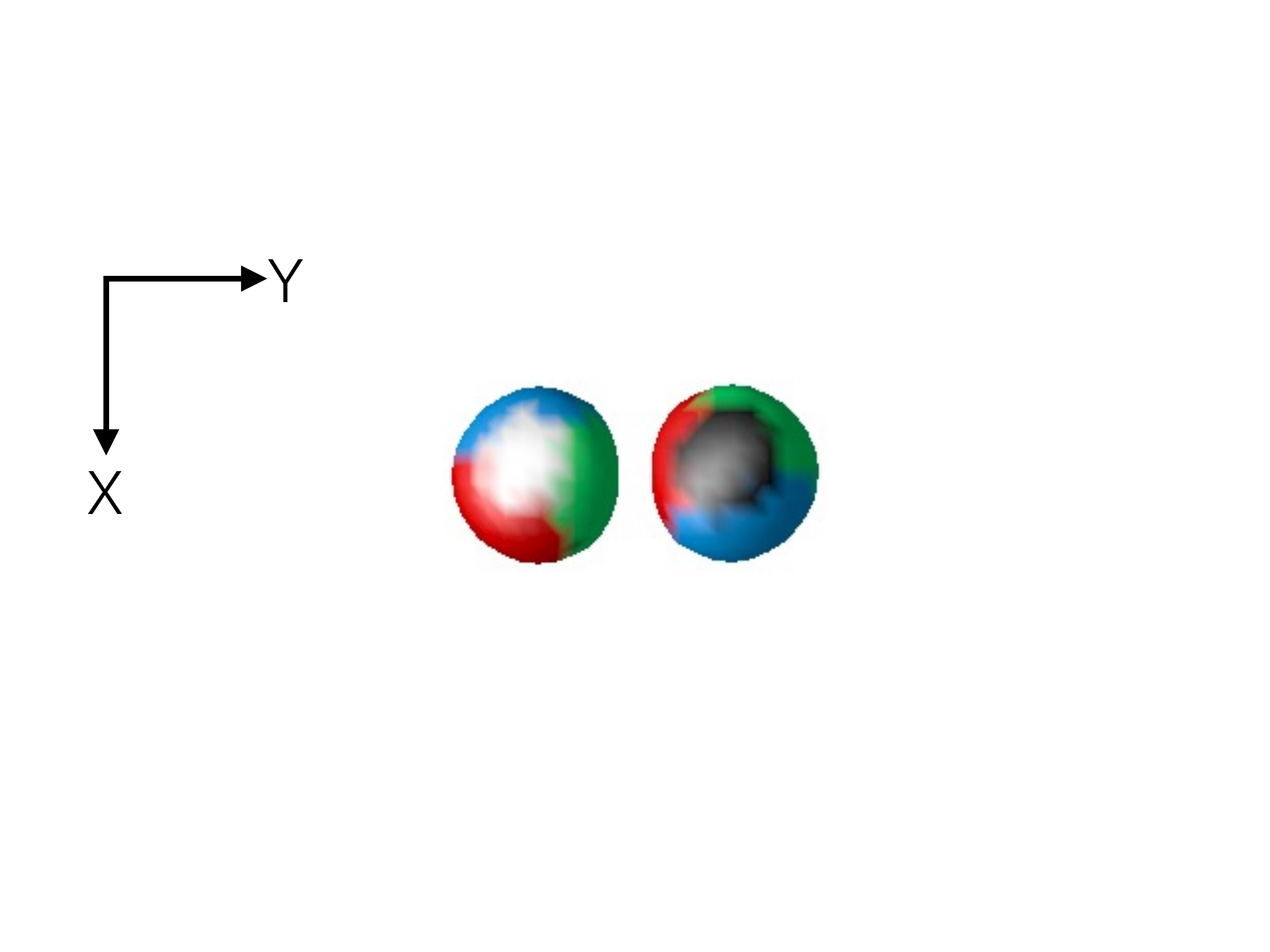}
               \caption*{ $t=t_0$}
   
       \end{subfigure}%
        ~ 
        \begin{subfigure}[b]{0.3\textwidth}
                \centering \includegraphics[width=\textwidth]{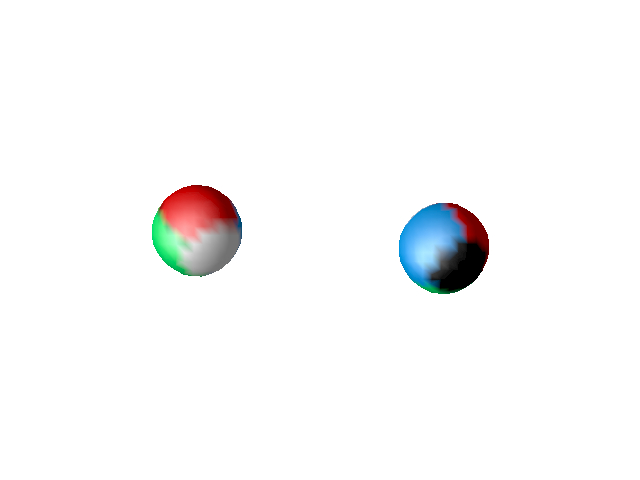}
                \caption*{$t=t_1$}
            
        \end{subfigure}
        ~
          \begin{subfigure}[b]{0.3\textwidth}
                \centering \includegraphics[width=\textwidth]{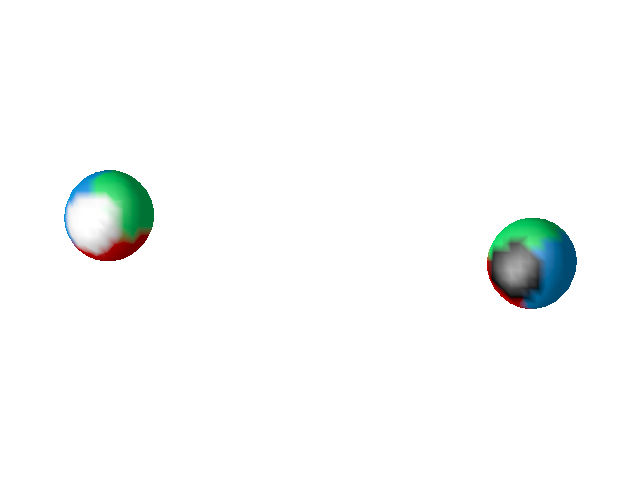}
                \caption*{$t=t_2$}
        
        \end{subfigure}

\caption{The repulsion of a \sk proton and a \sk neutron with zero
  initial speed. The dynamics is shown online.}
\label{np-v0}
\end{figure}

There is, however, a problem to be overcome. If we minimise the
energy by choosing initial speed $\mbox{v}=0$, and 
the initial relative orientations arbitrary, then averaged over time the 
Skyrmions repel. This is because the Skyrmions are spinning, and most of the
time the colours do not match, producing a repulsion. Only briefly do
the colours match, and the Skyrmions attract. Similar behaviour was
observed for two spinning baby Skyrmions in 2-dimensions
\cite{Piette:1994mh}. The repulsion of spinning Skyrmions initially at rest,
which is not a Coulomb effect, is shown in \fg \ref{np-v0}.

\begin{figure}[H]
       \centering
           
          \begin{subfigure}[b]{0.22\textwidth}
               \centering
       \begin{overpic}[width=\textwidth]{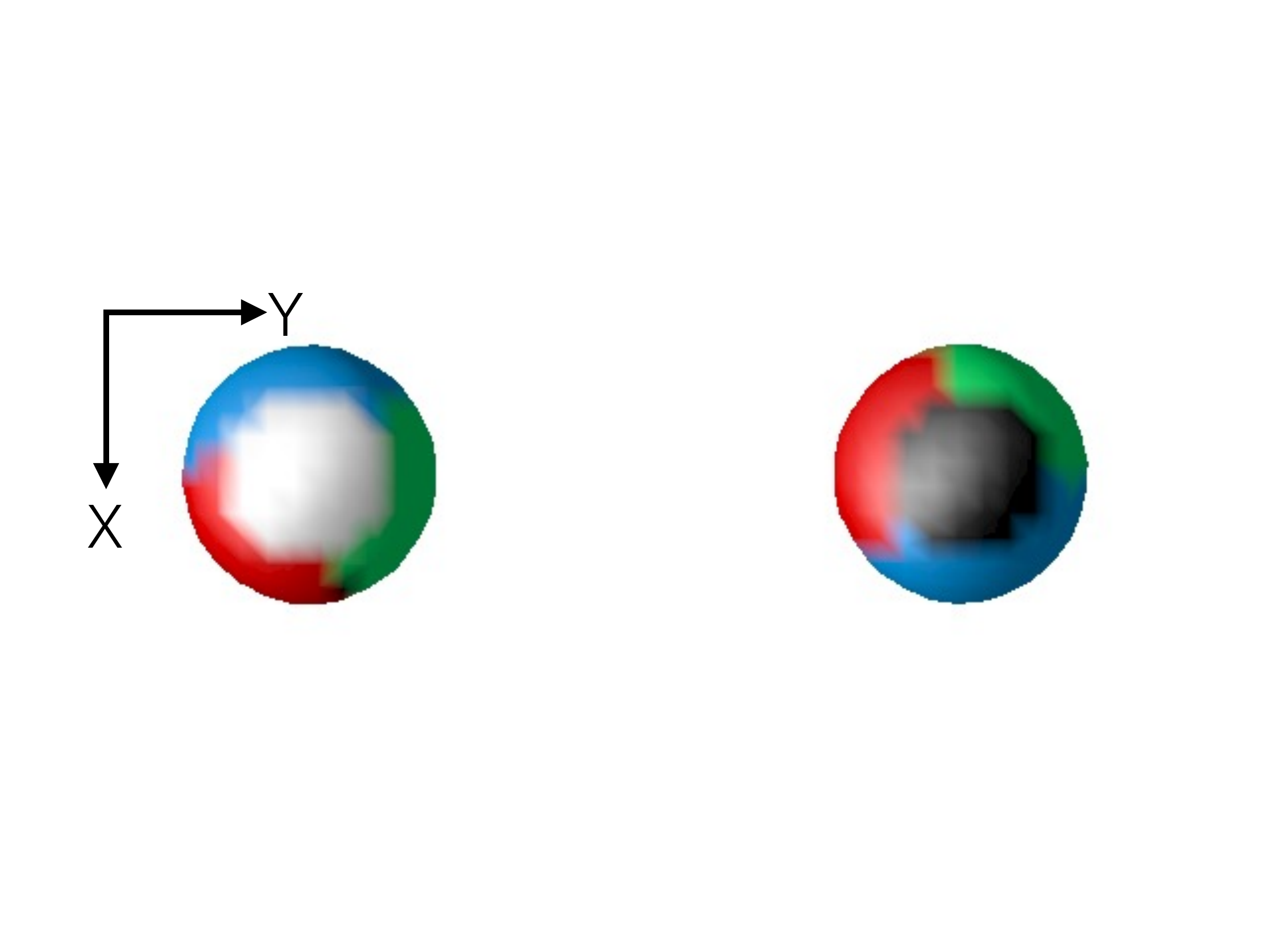}\put (10,55) {Top}
\end{overpic}
               \caption*{$t=t_0$}
   
       \end{subfigure}%
        ~ 
        \begin{subfigure}[b]{0.22\textwidth}
                \centering \includegraphics[width=\textwidth]{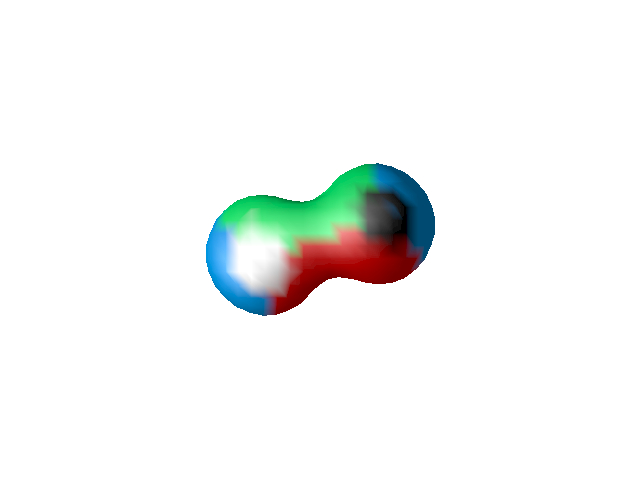}
                \caption*{$t=t_1$}
            
        \end{subfigure}
        ~
          \begin{subfigure}[b]{0.22\textwidth}
                \centering \includegraphics[width=\textwidth]{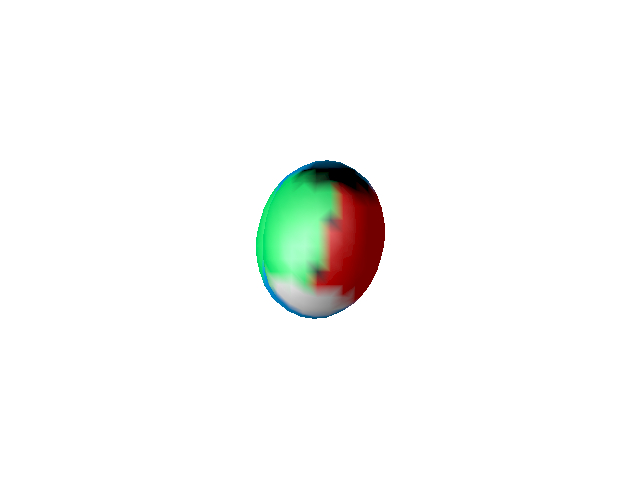}
                \caption*{$t=t_2$}
        
        \end{subfigure}
         ~
          \begin{subfigure}[b]{0.22\textwidth}
                \centering \includegraphics[width=\textwidth]{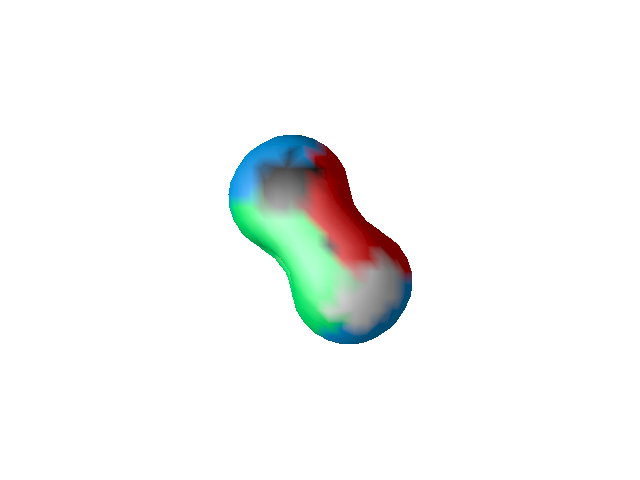}
                \caption*{$t=t_3$}
           
        \end{subfigure}
        \\
       \begin{subfigure}[b]{0.22\textwidth}
               \centering
         \begin{overpic}[width=\textwidth]{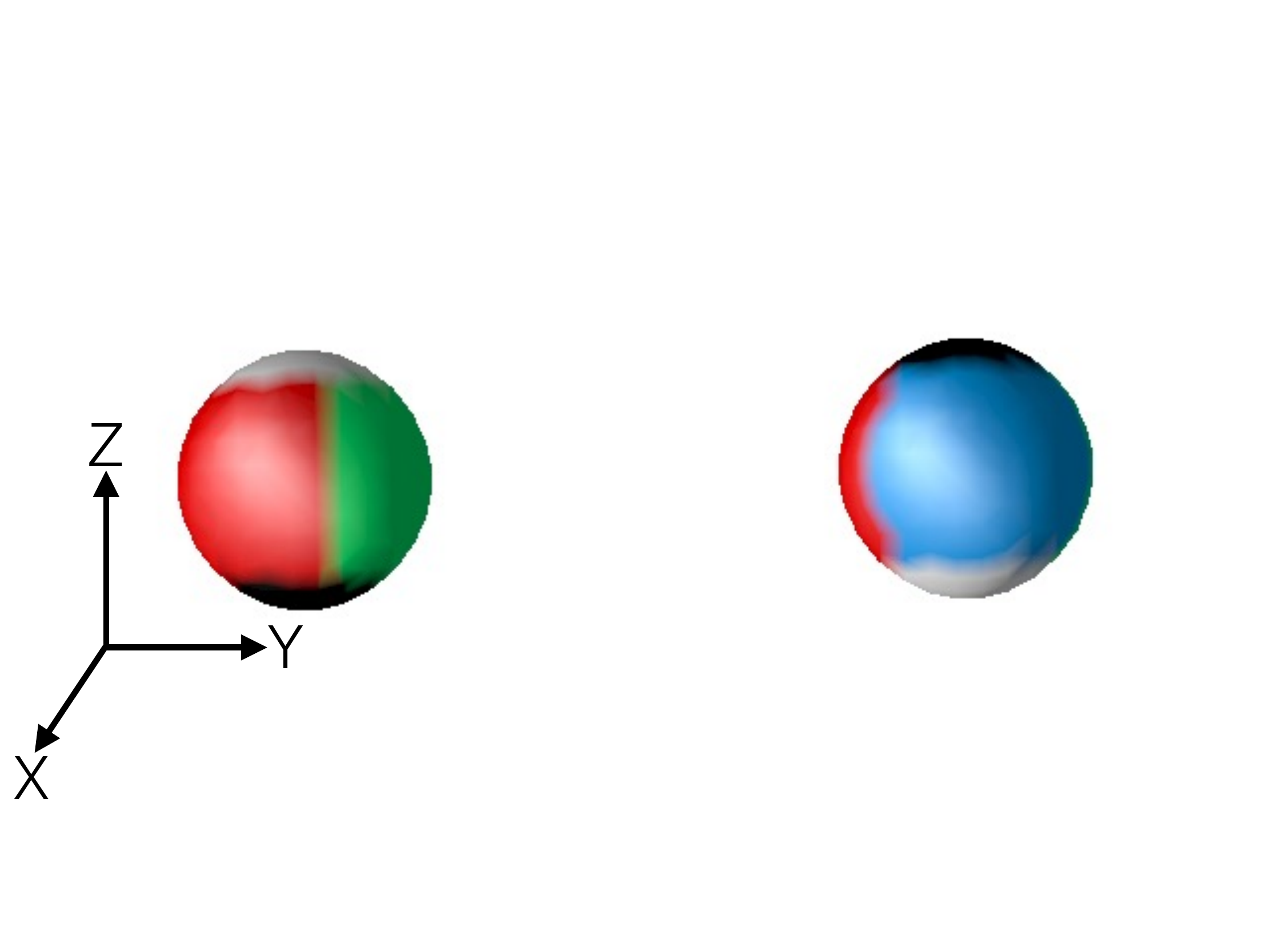}\put (10,55) {Side}
\end{overpic}
               \caption*{$t=t_0$}
         
       \end{subfigure}%
        \! 
        \begin{subfigure}[b]{0.22\textwidth}
                \centering \includegraphics[width=\textwidth]{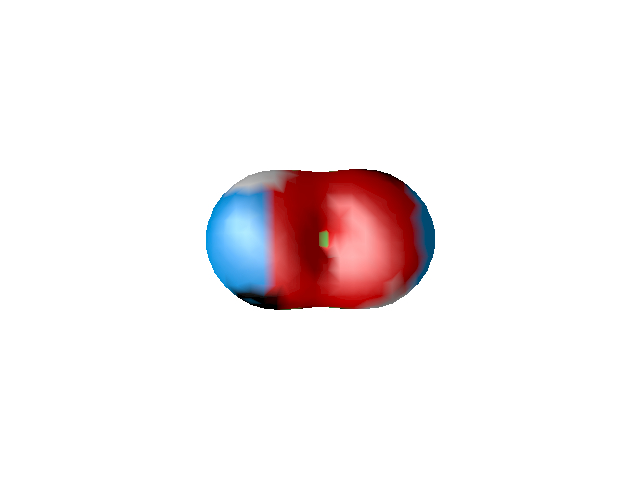}
                \caption*{$t=t_1$}
       
        \end{subfigure}
        \!
          \begin{subfigure}[b]{0.22\textwidth}
                \centering \includegraphics[width=\textwidth]{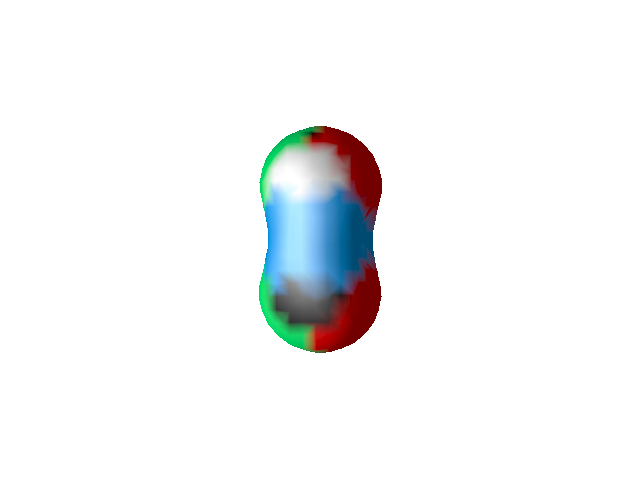}
                \caption*{$t=t_2$}
              
        \end{subfigure}
         \!
          \begin{subfigure}[b]{0.22\textwidth}
                \centering \includegraphics[width=\textwidth]{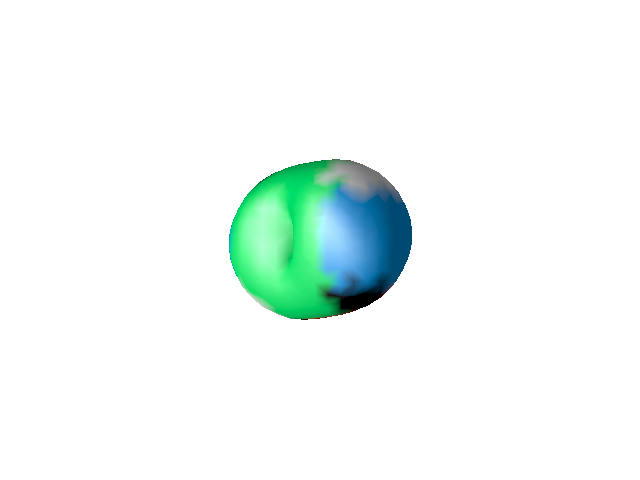}
                \caption*{$t=t_3$}
                
        \end{subfigure}
    
\caption{Formation of a deuteron bound state in a spinning Skyrmion
collision with $\bJ=1$, $\bI=0$. This is shown dynamically online.}
\label{deuteron}
\end{figure}

Therefore, to model deuteron formation we need to collide the \sks at a positive
speed, and we choose $\mbox{v}=0.2$. There is now enough kinetic energy 
to overcome the repulsive barrier. The dynamic $B=2$ \sk that forms is 
stable even with this extra energy. The collision along the $y$-axis
of two Skyrmions spinning about the $z$-direction is shown in \fg
\ref{deuteron}, from two viewpoints.

As expected \cite{Forest:1996kp}, the $B=2$ torus
forms in the $(y,z)$-plane and spins about the $z$-axis, which is an 
axis orthogonal to the symmetry axis of the torus. Therefore there is
no projection of spin or isospin along the symmetry axis, and the constraint 
$L_3+2K_3=0$ is satisfied.

\begin{figure}[htb] 
\centering
\includegraphics[width=\textwidth]{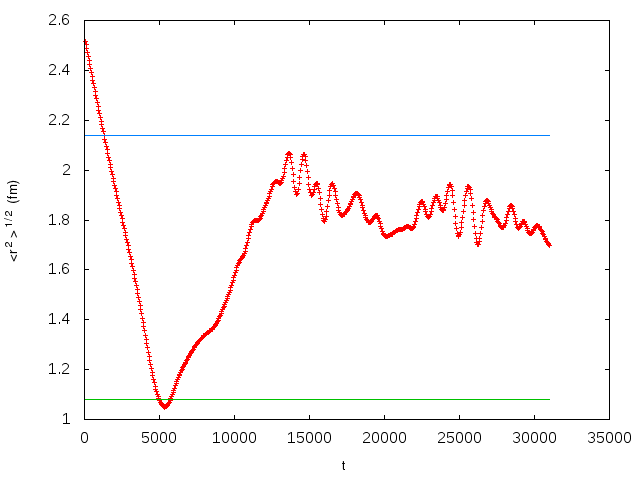}
\caption{\sk matter radius in the classical simulation of a 
neutron-proton collision, as a function of time $t$. The 
upper line is the experimental deuteron matter radius and 
the lower line is the matter radius of the static $B=2$ toroidal \sk.}
\label{matter_rad}
\end{figure}

The important feature of this collision is that the $B=1$ Skyrmions do not
scatter, but merge into a long-lived spinning $B=2$ Skyrmion, with 
superimposed oscillations. The colours are not independently spinning. When the 
oscillation is at its maximal amplitude, two $B=1$ \sks can be 
identified. They represent two nucleons with parallel spins. 
This classical bound state models the physical, quantised 
deuteron \cite{Braaten:1988cc}, with the correct spin, and zero isospin. 
The oscillation is similar to that of the $B=2$ \sk in its quantised 
ground state obtained using the instanton ansatz \cite{Leese:1994hb}, but
the amplitude of oscillation is somewhat greater, and the energy
somewhat larger. The classical oscillation produces an oscillating 
matter radius, which is shown in \fg \ref{matter_rad} as a function of time.
The time-averaged matter radius is comparable with the experimental 
value for the deuteron.

\begin{figure}[H]
       \centering
       \begin{subfigure}[b]{0.3\textwidth}
               \centering
    \includegraphics[width=\textwidth]{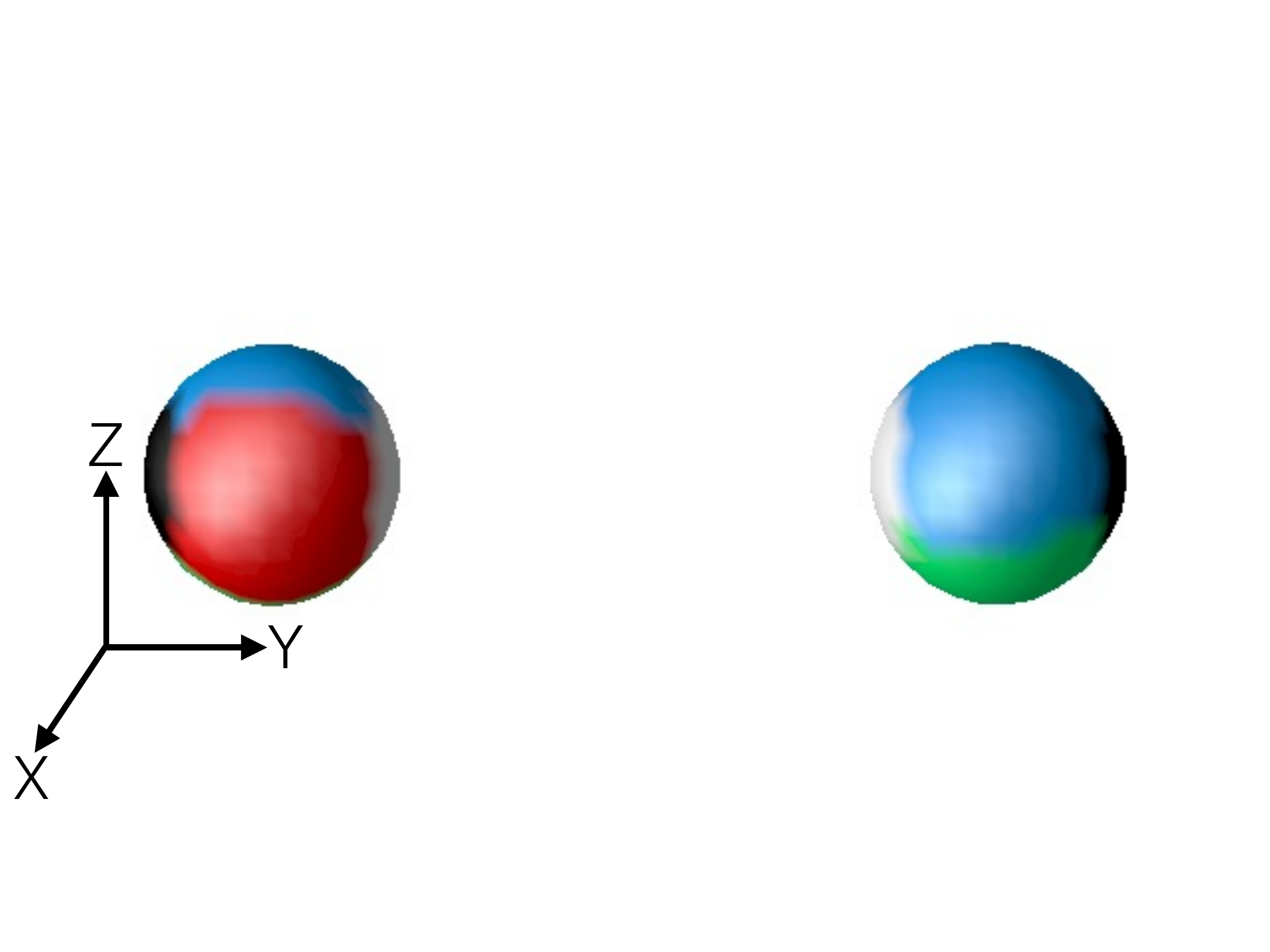}
               \caption*{$t=t_0$} 
       \end{subfigure}%
        ~ 
        \begin{subfigure}[b]{0.3\textwidth}
                \centering \includegraphics[width=\textwidth]{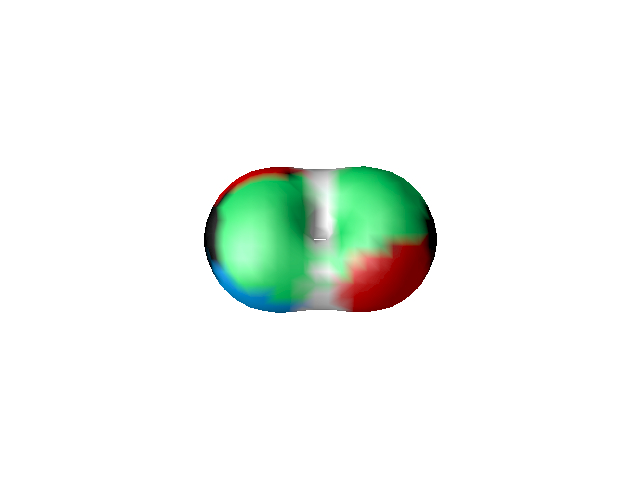}
                \caption*{$t=t_1$}
            
        \end{subfigure}
        ~
          \begin{subfigure}[b]{0.3\textwidth}
                \centering \includegraphics[width=\textwidth]{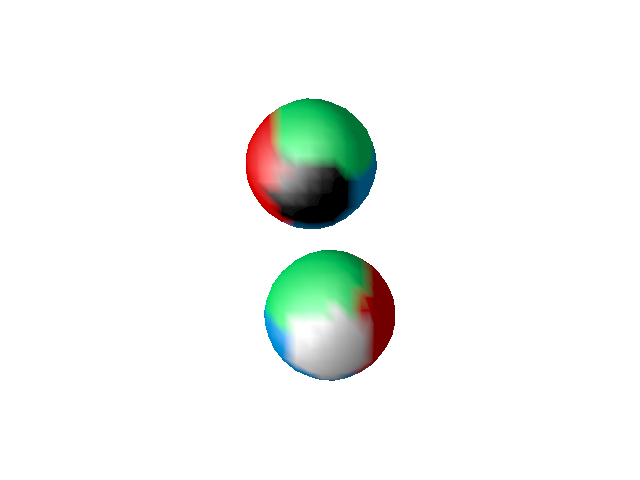}
                \caption*{$t=t_2$}
           
        \end{subfigure}
        \caption{Head-on collision of linearly polarised proton and neutron
       with $\bJ=1$, $\bI=0$. Note: at $t=t_1$ the torus is spinning
       about its black-black axis. The dynamics is shown online.}
        \label{l0-np-scattering}
\end{figure}

We have also attempted to simulate deuteron formation by colliding
\sks spinning as a neutron and proton, both with linear polarisation. 
This is shown in \fg \ref{l0-np-scattering}. The process does 
not produce a classical bound state, even though the spin and isospin 
are correct for a deuteron. The initial state can 
be understood as \sks approaching in the attractive channel with two
white faces being closest. The \sks scatter at right angles 
and the initial spin is converted to orbital angular momentum, the 
\sks separating with some non-zero impact
parameter. They also spin slowly about a coloured axis as they separate -- 
this is inferred from a scattering of \sks with larger initial spin. Skyrmions spinning like
this should be interpreted as a superposition of a neutron and a proton. 

Although the torus forms briefly in the linearly polarised case, it 
breaks apart. This is because \sks in the attractive channel 
accelerate towards each other and the excess 
energy quickly breaks the torus up. In the transversely polarised case the 
excess energy is less and goes into the vibrational mode. This shows 
that polarisation has implications for proton-neutron fusion.

\subsection{Excited Dibaryon}

Recently an excited proton-neutron state has been discovered 
\cite{Adlarson:2014pxj}, called the dibaryon. It can be viewed 
as an excited spin 3 state of the spin 1 deuteron, with mass 
$M_{\rm Dib}\approx2380$ MeV and quantum numbers $\bJ=3$, $\bI=0$.
 
\begin{figure}[H]
       \centering
       \begin{subfigure}[b]{0.3\textwidth}
               \centering
    \begin{overpic}[width=\textwidth]{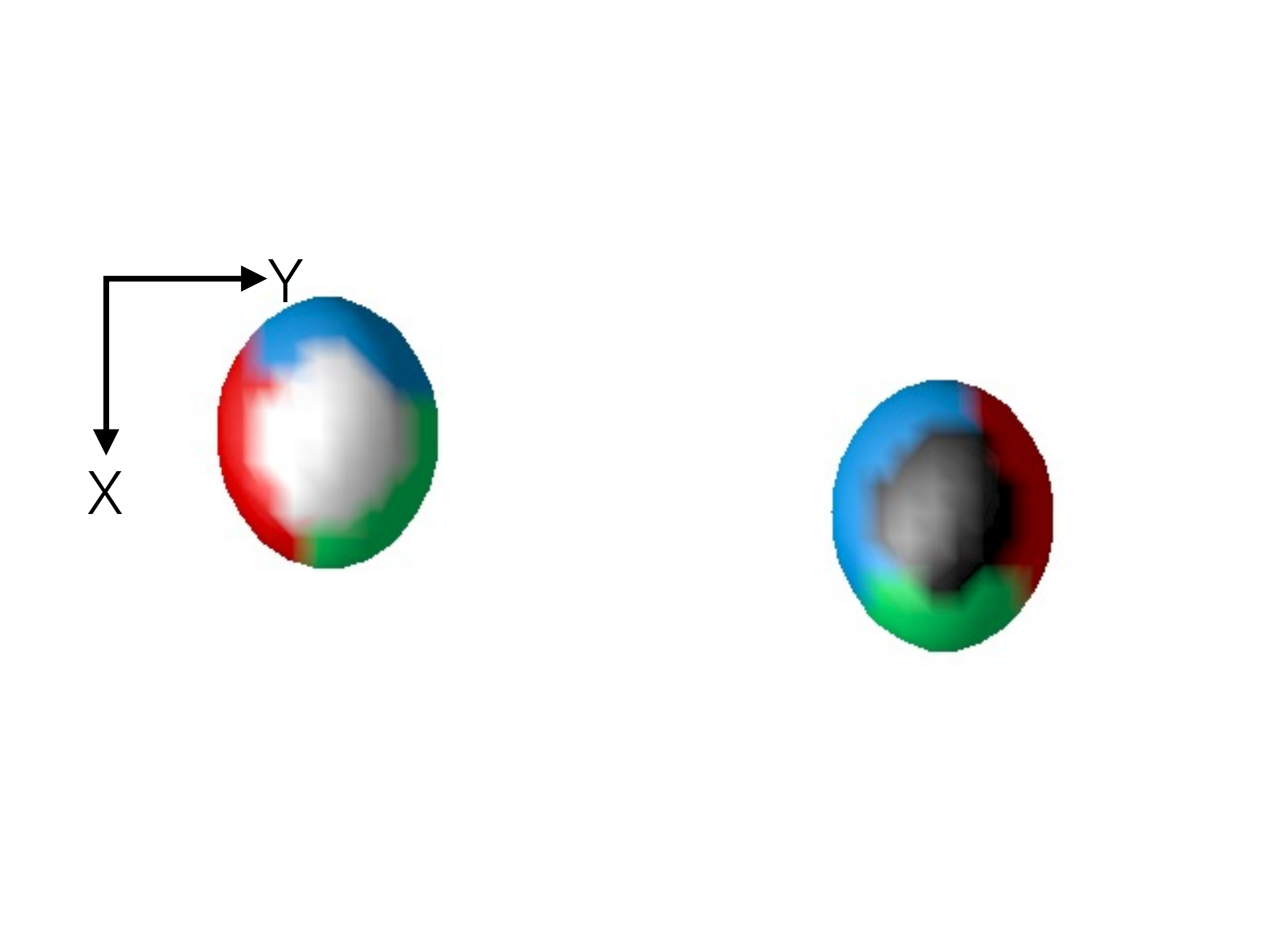}\put (10,60) {Top}
\end{overpic}
               \caption*{$t=t_0$}
             
       \end{subfigure}%
        ~ 
        \begin{subfigure}[b]{0.3\textwidth}
                \centering \includegraphics[width=\textwidth]{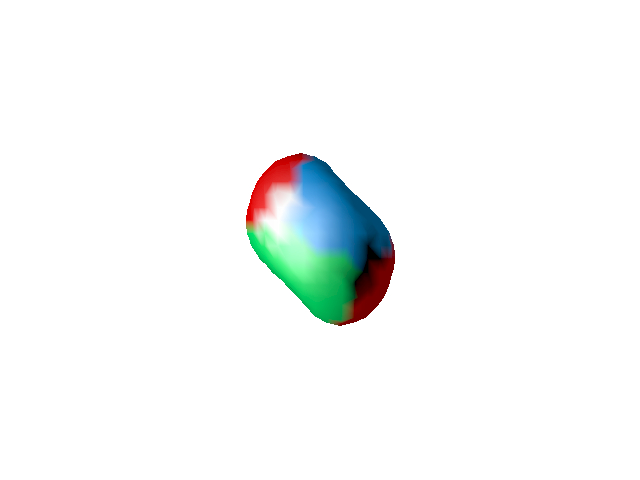}
                \caption*{$t=t_1$}
            
        \end{subfigure}
        ~
          \begin{subfigure}[b]{0.3\textwidth}
                \centering \includegraphics[width=\textwidth]{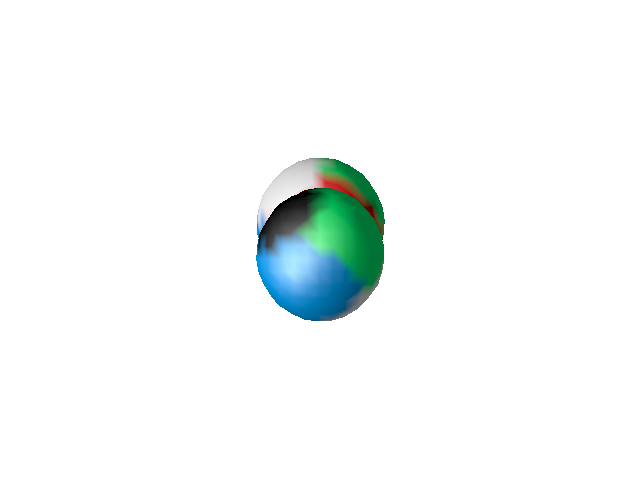}
                \caption*{$t=t_2$}
           
        \end{subfigure}
        \\
               \begin{subfigure}[b]{0.3\textwidth}
                \centering 
                  \begin{overpic}[width=\textwidth]{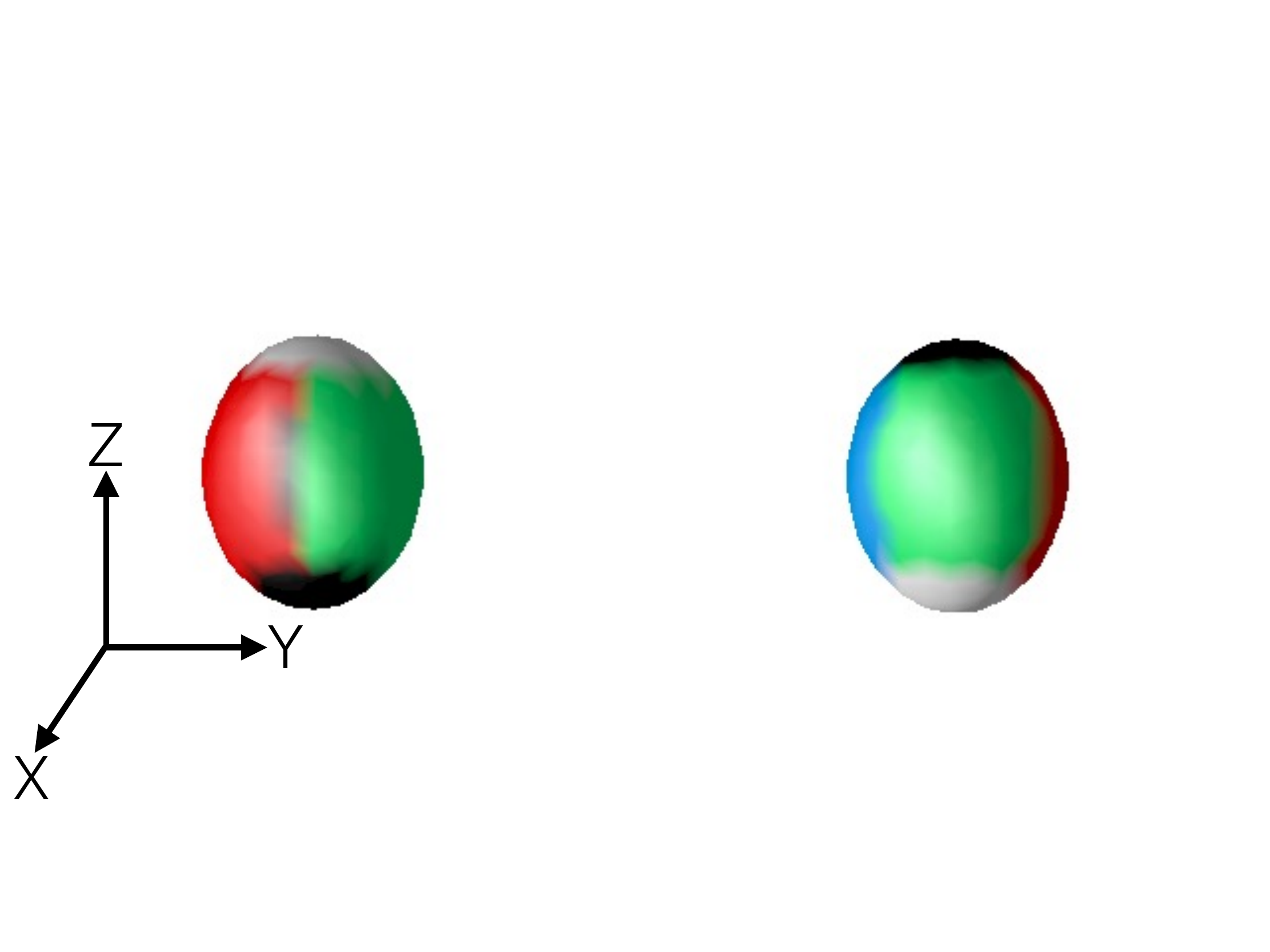}\put (10,55) {Side}
\end{overpic}
                \caption*{$t=t_0$}
                
        \end{subfigure} 
        ~
               \begin{subfigure}[b]{0.3\textwidth}
                \centering \includegraphics[width=\textwidth]{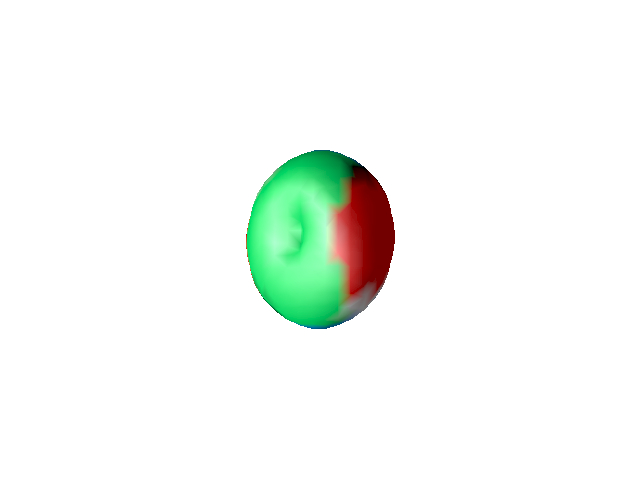}
                \caption*{$t=t_1$}
     
        \end{subfigure}
        ~
          \begin{subfigure}[b]{0.3\textwidth}
                \centering \includegraphics[width=\textwidth]{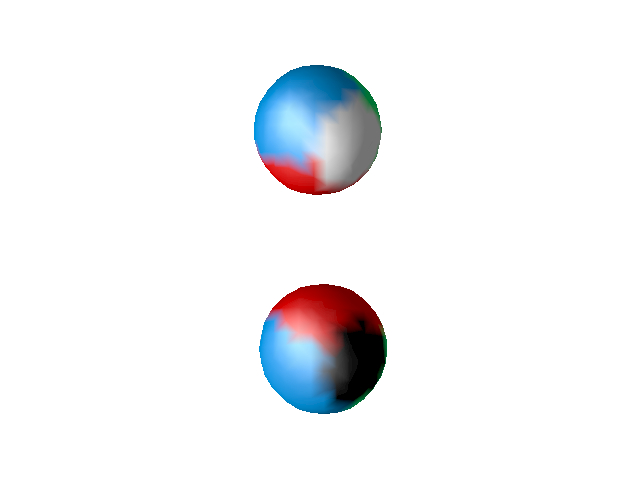}
                \caption*{$t=t_2$}
           
        \end{subfigure}
        \caption{Skyrmion scattering modelling proton-neutron
scattering with $\bJ=3$, $\bI=0$ at the energy of the dibaryon 
resonance. The orbital angular momentum is $l=2$, requiring an impact 
parameter $b=0.32$. Note: at $t=t_2$ the \sks are spinning about the 
red-cyan axis. The dynamical scattering is shown online.}
        \label{L2-scattering}
\end{figure}

To create the \sk analogue requires colliding two \sks each 
with kinetic energy $T = \frac{1}{2}(M_{\rm Dib}-2M_{\rm N})$; 
this is achieved with initial, relativistic speeds of
$\mbox{v}=0.6$. The $\bJ=3$, $\bI=0$ state is replicated by
colliding a \sk spinning as a neutron and an oppositely orientated \sk
spinning as a proton, with the spins parallel and with orbital angular 
momentum $l=2$, such that $J_3=\frac{1}{2}+\frac{1}{2}+2$. The orbital 
angular momentum is achieved with an impact parameter $b=0.32$ at this 
speed. This impact parameter is the same order of magnitude 
as half the spatial width of the $B=2$ toroidal Skyrmion. The
scattering is shown in \fg \ref{L2-scattering}, with the
initial motion parallel to the $y$-axis, and the proton and neutron
polarisations parallel to the $z$-axis. The dibaryon appears briefly 
in the form of the $B=2$ toroidal \sk spinning with $\bJ=3$ about 
the $z$-axis. The spin axis is at right angles to the torus's 
symmetry axis, as for the deuteron.  

There is a small amount of excess energy and as a result the $B=2$
Skyrmion almost breaks up into two rapidly spinning $B=1$ Skyrmions
moving back-to-back along the $z$-axis. These have close to spin
$\frac{3}{2}$ each, but rapidly lose energy, presumably by pion radiation,
although this is hard to see in the simulation. The spinning Skyrmions
stop and return to form a spinning $B=2$ Skyrmion again. This
replicates the experimental result that two pions are emitted, and a
deuteron remains.

The transient production of spin $\frac{3}{2}$ particles matches other
theoretical models of the dibaryon resonance involving two deltas with
spin $\frac{3}{2}$ \cite{GalGar}. Our simulation shows that (in the 
proton-neutron centre-of-mass frame) the deltas are emitted
transversely to the line of collision, but parallel to the proton and 
neutron polarisation axes. Each delta is linearly polarised, and because its spin axis 
is an axis perpendicular to the white-black axis, it is in a superposition of states 
with different charges. This superposition is not surprising, because the average charge
emerging along the positive or negative $z$-axis is half the proton charge.

\begin{figure}[H]
       \centering
       \begin{subfigure}[b]{0.3\textwidth}
               \centering
      \begin{overpic}[width=\textwidth]{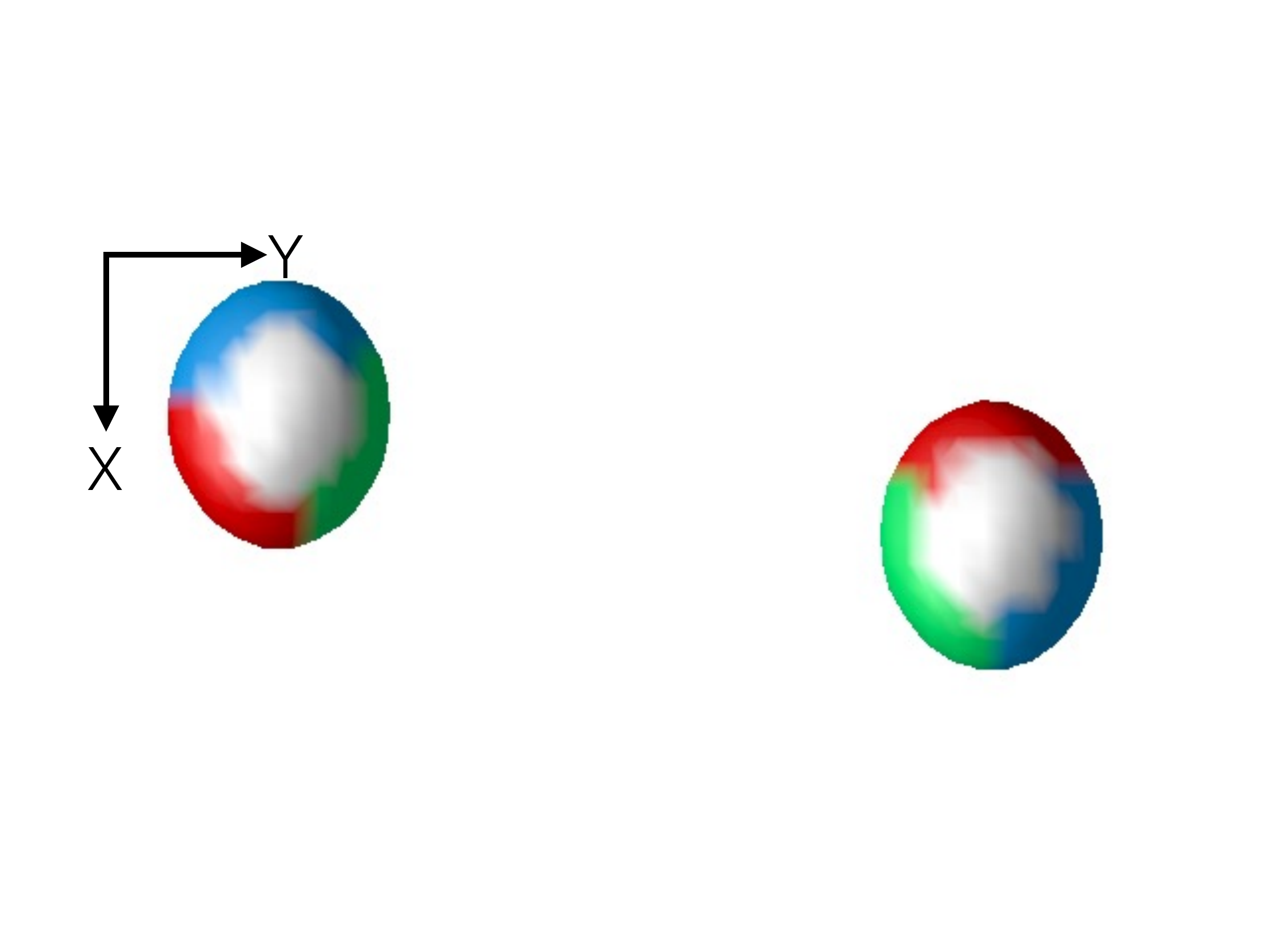}\put (10,60) {Top}
\end{overpic}
               \caption*{$t=t_0$}
              
       \end{subfigure}%
        ~ 
        \begin{subfigure}[b]{0.3\textwidth}
                \centering \includegraphics[width=\textwidth]{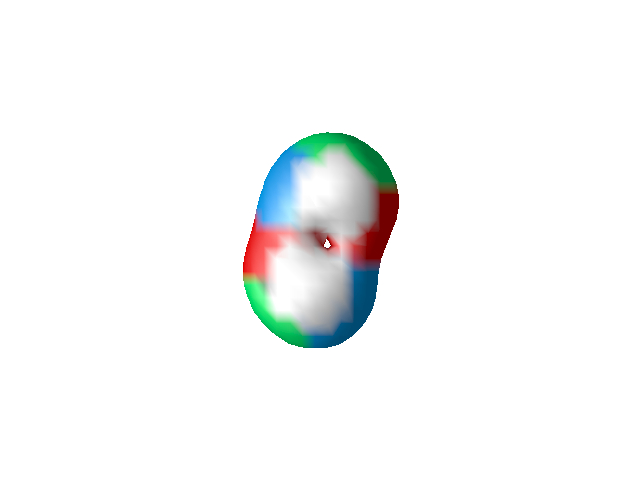}
                \caption*{$t=t_1$}
            
        \end{subfigure}
        ~
          \begin{subfigure}[b]{0.3\textwidth}
                \centering \includegraphics[width=\textwidth]{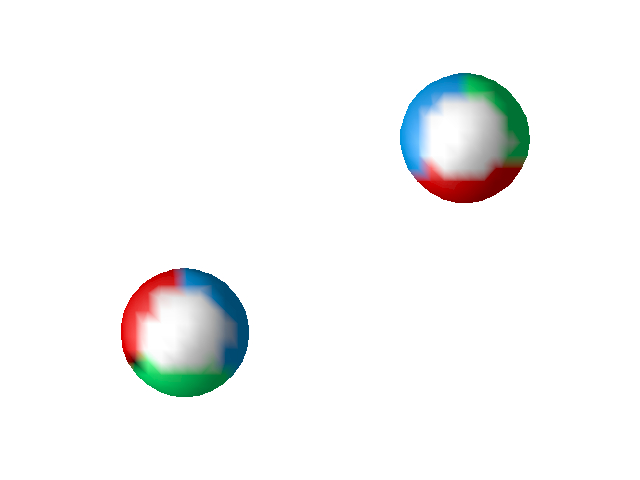}
                \caption*{$t=t_2$}
           
        \end{subfigure}
        \\
               \begin{subfigure}[b]{0.3\textwidth}
                \begin{overpic}[width=\textwidth]{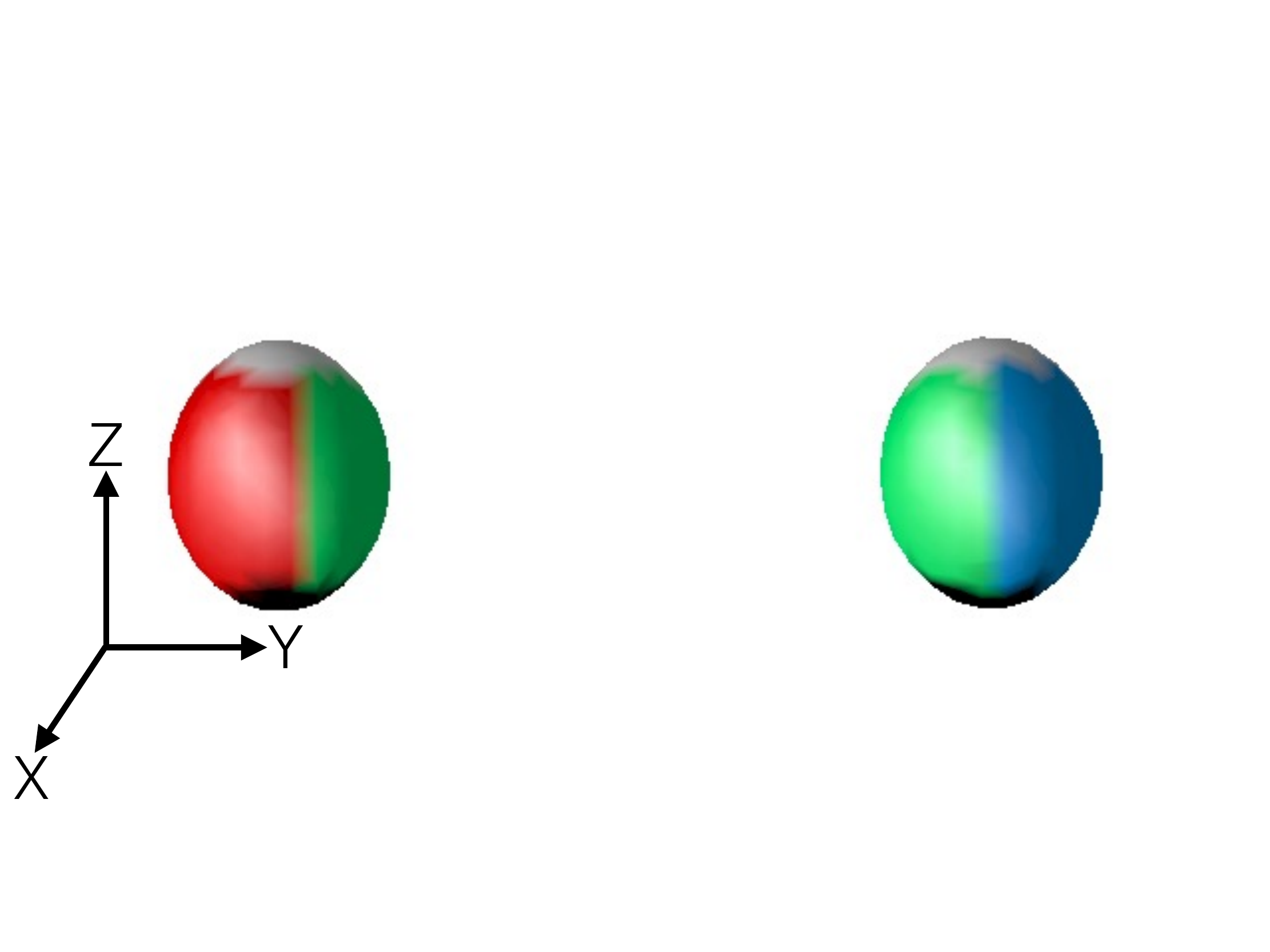}\put (10,55) {Side}
\end{overpic}
                \caption*{$t=t_0$}
                
        \end{subfigure} 
        ~
               \begin{subfigure}[b]{0.3\textwidth}
                \centering \includegraphics[width=\textwidth]{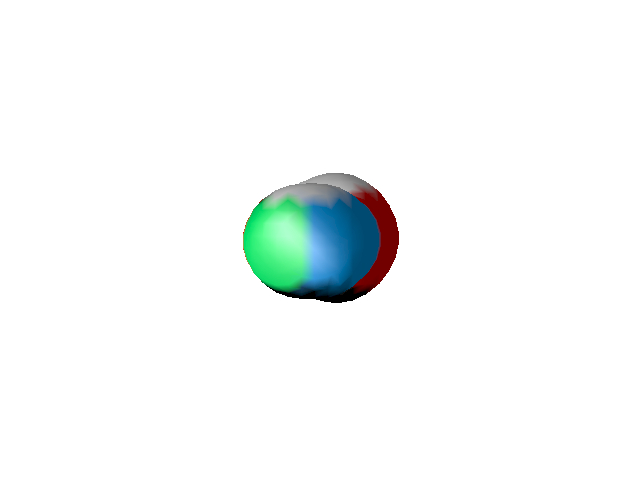}
                \caption*{$t=t_1$}
     
        \end{subfigure}
        ~
          \begin{subfigure}[b]{0.3\textwidth}
                \centering \includegraphics[width=\textwidth]{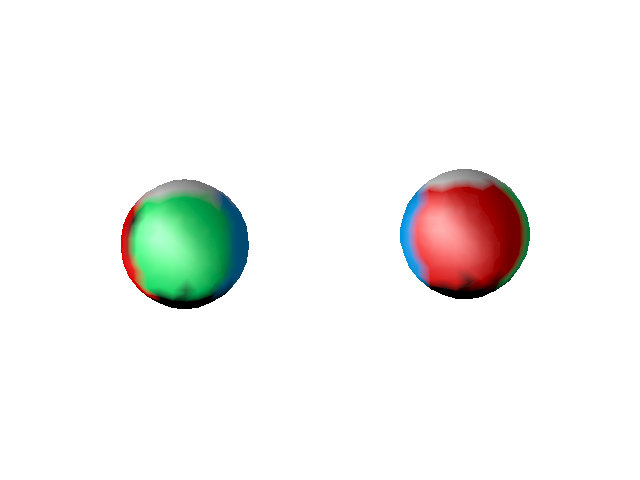}
                \caption*{$t=t_2$}
           
        \end{subfigure}
        \caption{Modelling dibaryon production in proton-neutron
          scattering, with orbital angular momentum $l=3$, and impact parameter $b=0.48$. This is
          shown dynamically online.}
        \label{L3-scattering}
\end{figure}

We have also attempted to produce a $\bJ=3$, $\bI=0$ dibaryon state by
colliding a proton \sk and a neutron \sk with the spin polarisations transverse 
but opposite, so that the spin angular momenta cancel. The orbital 
angular momentum is $l=3$, which requires the same initial velocities, 
but an impact parameter $b=0.48$. This scattering is
shown in \fg \ref{L3-scattering}.

Here, the toroidal \sk also appears but is not an
excited deuteron, because it forms in the plane normal to the initial
orbital angular momentum vector. In this case, one might anticipate the
torus to be spinning about its symmetry axis, but it would then have
non-zero isospin, which is not allowed. On closer inspection one 
observes that the $B=2$ torus is not spinning, but each $B=1$ \sk can 
be interpreted as a wave propagating along the outer edge, such that 
the colours do not rotate. This curious dynamical configuration is not
a dibaryon. 

In summary, dibaryon formation can be observed in
classical Skyrmion scattering, but only if the initial polarisations
of the proton and neutron are transverse and parallel.

\section{Conclusion}

In this work we have classically scattered spinning \sks to model 
nucleon-nucleon scattering. The initial spin angular momenta are fixed to be 
$\frac{1}{2} \hbar$. Interestingly, we have found short-lived states 
modelling the dineutron/diproton and the dibaryon. We have also
found that in a collision with the quantum numbers of the deuteron, 
the Skyrmions do not scatter, but merge into a spinning and oscillating 
form of the $B=2$ toroidal Skyrmion. So, even without quantisation, the 
Skyrme model usefully captures important physical features of 
two-nucleon dynamics. This is gratifying but a little surprising, 
especially when it is recalled that our long-lived deuteron state was made 
from two \sks colliding with a larger kinetic energy than that needed to
produce the short-lived dineutron/diproton.  

Our most valuable observations concern the polarisations. The classically
spinning Skyrmions must be polarised along some axis, so we inevitably
have complete polarisation information for the nucleons before and
after scattering. Such complete information is hard to obtain
experimentally, and also difficult to determine in more traditional
nucleon-nucleon potential models. Classically we also fix the impact
parameter. Several of our simulations lead to right angle scattering 
of the Skyrmions. Experimentally one
cannot fix the impact parameter, but one can focus on scattering
events where there is right angle scattering (in the centre-of-mass
frame), and can hope that these are modelled by our 
collisions. For such events, our Skyrmion simulations make strong
predictions for the polarisations. 

We have observed that in a $2$-proton or $2$-neutron head-on 
collision, with total angular momentum zero, if the particles are transversely
polarised initially then they are transversely polarised finally.
If they are linearly polarised initially they are
linearly polarised finally. In both cases the polarisation
axes rotate through a right angle. Moreover, there is a fixed plane 
containing the initial and final momenta and also the polarisations. 
The sense of the polarisations is determined by conservation of 
${\bf p} \times {\bf s}$ for each particle in the transverse 
polarisation case, and conservation of ${\bf p} \cdot {\bf s}$ in the linear 
polarisation case. It would be very interesting if something similar 
was observed experimentally.

In our simulations of a proton-neutron collision leading to the 
dibaryon, with total spin 3, we have seen the transient production of 
two linearly polarised deltas each with spin close to $\frac{3}{2}$. 

We intend to extend our numerical simulations to model nucleon-deuteron and
deuteron-deuteron collisions, possibly forming Helium-3, Tritium or
Helium-4 bound states. The physical spins must be correctly modelled 
in the initial data, and the results will differ from those obtained with 
non-spinning Skyrmions.

\vspace{-0.5cm}
\section{Acknowledgments}
\vspace{-0.3cm}
DF would like to thank Prof. David Jenkins, Dr. Steffen Krusch and 
Dr. Mareike Haberichter for useful conversations. DF acknowledges the Leverhulme Trust 
Program Grant: Scientific Properties Of Complex Knots.

\end{document}